%% file: radialdraft.tex
\documentclass[12pt]{article}
\pdfoutput=1
\usepackage{graphicx}

\usepackage[T1]{fontenc}
\usepackage[utf8]{inputenc}
\usepackage{ytableau}
\usepackage{hyperref}
\usepackage{amsmath}
\usepackage{amssymb}
\usepackage[all]{xy}
\usepackage{xspace}
\usepackage{braket}
\usepackage{multirow}
\usepackage{subfig}
\usepackage[percent]{overpic}

\newcommand{\hj}{{\hat \jmath}}
\newcommand{\hi}{{\hat \imath}}

\newcommand{\ii}{\mathrm{i}}

\newcommand{\hQ}{{\hat Q}}
\newcommand{\hR}{{\hat R}}
\newcommand{\hM}{{\hat M}}

\newcommand{\hOcal}{{\hat \Ocal}}
\newcommand{\hHcal}{{\hat \Hcal}}
\newcommand{\hP}{{\hat P}}

\newcommand{\hD}{{\hat \D}}

\newcommand{\hl}{{\hat l}}
\newcommand{\hr}{{\hat r}}
\newcommand{\heta}{{\hat \eta}}
\newcommand{\hG}{{\hat G}}
\newcommand{\hg}{{\hat g}}

\newcommand{\hDcal}{{\hat \Dcal}}

\newcommand{\hrho}{{\hat \rho}}
\newcommand{\hrhobar}{{\bar{\hat \rho}}}

\newcommand{\pdot}{{\, \bullet \,}}
\newcommand{\tdot}{{\, \circ\, }}
\newcommand{\Pp}{\Pi_{\bullet }}

\newcommand{\Pt}{\Pi_{\circ }}
\newcommand{\pp}{\pi_{\bullet }}
\newcommand{\pt}{\pi_{\circ}}

\newcommand{\gbullet}{{\bullet}}
\newcommand{\wbullet}{{\circ}}
\newcommand{\dd }{\mathtt{d}}
\newcommand{\hO }{{\hat \Ocal}}

\def\beq{\begin{equation}} 
\def\eeq{\end{equation}} 
\newcommand{\be}{\begin{equation}}
\newcommand{\ee}{\end{equation}} 
\newcommand{\ba}{\begin{eqnarray}}
\newcommand{\ea}{\end{eqnarray}}

\newcommand{\Acal}{{\mathcal A}}

\newcommand{\Ccal}{{\mathcal C}}

\newcommand{\Dcal}{{\mathcal D}}

\newcommand{\Hcal}{{\mathcal H}}

\newcommand{\Ocal}{{\mathcal O}}

\newcommand{\Vcal}{{\mathcal V}}
\newcommand{\Wcal}{{\mathcal W}}

\let\a=\alpha   \let\d=\delta 
    
\let\l=\lambda \let\m=\mu \let\n=\nu  \let\r=\rho
 \let\t=\tau   

  \let\D=\Delta

\newcommand{\I}{\textrm{I}\xspace}
\newcommand{\II}{\textrm{II}\xspace}
\newcommand{\III}{\textrm{III}\xspace}

\DeclareFontFamily{OMX}{MnSymbolE}{}
\DeclareSymbolFont{largesymbolsX}{OMX}{MnSymbolE}{m}{n}
\DeclareFontShape{OMX}{MnSymbolE}{m}{n}{
    <-6>  MnSymbolE5
   <6-7>  MnSymbolE6
   <7-8>  MnSymbolE7
   <8-9>  MnSymbolE8
   <9-10> MnSymbolE9
  <10-12> MnSymbolE10
  <12->   MnSymbolE12}{}

\DeclareMathSymbol{\downbrace}    {\mathord}{largesymbolsX}{'251}
\DeclareMathSymbol{\downbraceg}   {\mathord}{largesymbolsX}{'252}
\DeclareMathSymbol{\downbracegg}  {\mathord}{largesymbolsX}{'253}
\DeclareMathSymbol{\downbraceggg} {\mathord}{largesymbolsX}{'254}
\DeclareMathSymbol{\downbracegggg}{\mathord}{largesymbolsX}{'255}
\DeclareMathSymbol{\upbrace}      {\mathord}{largesymbolsX}{'256}
\DeclareMathSymbol{\upbraceg}     {\mathord}{largesymbolsX}{'257}
\DeclareMathSymbol{\upbracegg}    {\mathord}{largesymbolsX}{'260}
\DeclareMathSymbol{\upbraceggg}   {\mathord}{largesymbolsX}{'261}
\DeclareMathSymbol{\upbracegggg}  {\mathord}{largesymbolsX}{'262}
\DeclareMathSymbol{\braceld}      {\mathord}{largesymbolsX}{'263}
\DeclareMathSymbol{\bracelu}      {\mathord}{largesymbolsX}{'264}
\DeclareMathSymbol{\bracerd}      {\mathord}{largesymbolsX}{'265}
\DeclareMathSymbol{\braceru}      {\mathord}{largesymbolsX}{'266}
\DeclareMathSymbol{\bracemd}      {\mathord}{largesymbolsX}{'267}
\DeclareMathSymbol{\bracemu}      {\mathord}{largesymbolsX}{'270}
\DeclareMathSymbol{\bracemid}     {\mathord}{largesymbolsX}{'271}

\makeatletter
\def\horiz@expandable#1#2#3#4#5#6#7#8{%
  \@mathmeasure\z@#7{#8}%
  \@tempdima=\wd\z@
  \@mathmeasure\z@#7{#1}%
  \ifdim\noexpand\wd\z@>\@tempdima
    $\m@th#7#1$%
  \else
    \@mathmeasure\z@#7{#2}%
    \ifdim\noexpand\wd\z@>\@tempdima
      $\m@th#7#2$%
    \else
      \@mathmeasure\z@#7{#3}%
      \ifdim\noexpand\wd\z@>\@tempdima
        $\m@th#7#3$%
      \else
        \@mathmeasure\z@#7{#4}%
        \ifdim\noexpand\wd\z@>\@tempdima
          $\m@th#7#4$%
        \else
          \@mathmeasure\z@#7{#5}%
          \ifdim\noexpand\wd\z@>\@tempdima
            $\m@th#7#5$%
          \else
           #6#7%
          \fi
        \fi
      \fi
    \fi
  \fi}

\def\overbrace@expandable#1#2#3{\vbox{\m@th\ialign{##\crcr
  #1#2{#3}\crcr\noalign{\kern2\p@\nointerlineskip}%
  $\m@th\hfil#2#3\hfil$\crcr}}}
\def\underbrace@expandable#1#2#3{\vtop{\m@th\ialign{##\crcr
  $\m@th\hfil#2#3\hfil$\crcr
  \noalign{\kern2\p@\nointerlineskip}%
  #1#2{#3}\crcr}}}

\def\overbrace@#1#2#3{\vbox{\m@th\ialign{##\crcr
  #1#2\crcr\noalign{\kern2\p@\nointerlineskip}%
  $\m@th\hfil#2#3\hfil$\crcr}}}
\def\underbrace@#1#2#3{\vtop{\m@th\ialign{##\crcr
  $\m@th\hfil#2#3\hfil$\crcr
  \noalign{\kern2\p@\nointerlineskip}%
  #1#2\crcr}}}

\def\bracefill@#1#2#3#4#5{$\m@th#5#1\leaders\hbox{$#4$}\hfill#2\leaders\hbox{$#4$}\hfill#3$}

\def\downbracefill@{\bracefill@\braceld\bracemd\bracerd\bracemid}
\def\upbracefill@{\bracefill@\bracelu\bracemu\braceru\bracemid}

\def\upbrace@expandable{%
  \horiz@expandable
    \upbrace
    \upbraceg
    \upbracegg
    \upbraceggg
    \upbracegggg
    \upbracefill@}
\def\downbrace@expandable{%
  \horiz@expandable
    \downbrace
    \downbraceg
    \downbracegg
    \downbraceggg
    \downbracegggg
    \downbracefill@}

\DeclareRobustCommand{\overbrace}[1]{\mathop{\mathpalette{\overbrace@expandable\downbrace@expandable}{#1}}\limits}
\DeclareRobustCommand{\underbrace}[1]{\mathop{\mathpalette{\underbrace@expandable\upbrace@expandable}{#1}}\limits}
\makeatother

\begin{document}

\begin{titlepage}

\begin{center}
\vspace{1.5cm}

{\Large \bf Radial coordinates for defect CFTs}

\vspace{0.8cm}

{\bf Edoardo Lauria$^{1,4}$, Marco Meineri$^{2}$,  Emilio Trevisani$^{3,4}$}

\vspace{.5cm}
{
\small
{\it $^{1}$ Instituut voor Theoretische Fysica, KU Leuven, \\Celestijnenlaan 200D, B-3001 Leuven, Belgium \\
\it $^{2}$Institute of Physics, École Polytechnique Fédérale de Lausanne (EPFL), \\
 CH-1015 Lausanne, Switzerland \\
\it  $^3$Centro de Fisica do Porto,
 Departamento de Fisica e Astronomia,\\
Faculdade de Ci\^encias da Universidade do Porto,
Porto, Portugal
\\
Laboratoire de Physique Th\'eorique, \'Ecole Normale Sup\'erieure \& PSL Research University,
24 rue Lhomond, 75231 Paris Cedex 05, France
\\
\it $^{4}$ Perimeter Institute for Theoretical Physics \\
31 Caroline Street North, ON N2L 2Y5, Canada \\
}}
\end{center}

\vspace{1cm}

\begin{abstract}

We study the two-point function of local operators in the presence of a defect in a generic conformal field theory. We define two pairs of cross ratios, which are convenient in the analysis of the OPE in the bulk and defect channel respectively. The new coordinates have a simple geometric interpretation, which can be exploited to efficiently compute conformal blocks in a power expansion.  We illustrate this fact in the case of scalar external operators. We also elucidate the convergence properties of the bulk and defect OPE decompositions of the two-point function. In particular, we remark that the expansion of the two-point function in powers of the new cross ratios converges everywhere, a property not shared by the cross ratios customarily used in defect CFT. We comment on the crucial relevance of this fact for the numerical bootstrap.  

\end{abstract}

\bigskip

\end{titlepage}


\tableofcontents

\section{Introduction and summary}

Extended operators provide interesting probes of generic quantum field theories (QFT). The path-integral with the insertion of a defect describes the response of a theory to the presence of an impurity -- see \emph{e.g.} \cite{Billo:2013jda,Soderberg:2017oaa} -- its interaction with a boundary, an interface, or heavy source like a Wilson line -- see \cite{Cooke:2017qgm,Bianchi:2017ozk,Beccaria:2017rbe} for some recent work on the topic.
Defects are also a useful tool in more abstract constructions: for instance, they even capture information theoretic aspects of quantum field theory \cite{Calabrese:2004eu,Bianchi:2015liz,Balakrishnan:2017bjg}. The study of general properties of defects embedded in conformal field theories (CFTs), in particular, has a long history dating back at least to the pioneering work of Cardy on two dimensional CFTs \cite{Cardy:1984bb}. In higher dimensions, the successes of the conformal bootstrap of the four-point function of local operators \cite{arXiv:0807.0004} - see \cite{Simmons-Duffin:2016gjk} for an introduction and references - has encouraged similar explorations in the domain of defect CFT. Studies of the bootstrap constraints on defects in higher dimensions have appeared in recent years \cite{arXiv:1210.4258,arXiv:1502.07217,Gaiotto:2013nva,Liendo:2016ymz,Gliozzi:2016cmg}. Symmetry constraints on correlation functions of local operators with a defect have also been analyzed \cite{McAvity:1995zd,Billo:2016cpy}, and in \cite{Rastelli:2017ecj} the Mellin formalism was adapted to boundary CFTs. The present work fits in this program. We consider the simplest correlator which in the presence of a defect is subject to a crossing constraint, $\emph{i.e.}$ the two-point function of local operators. We define a new set of cross ratios, and we illustrate their convenience in the computation of the conformal blocks. Finally, we use the new coordinates to elucidate some aspects of the convergence of the OPE decomposition of the two-point function.

Before discussing the details, let us define the main player. A conformal defect is here taken to be a modification of the theory along a submanifold, which reduces the spacetime symmetry to the conformal transformations that preserve the submanifold. We also assume this operator to have a nontrivial overlap with the vacuum, so that its expectation value can be normalized to one.
We shall focus on the case of a spherical or flat defect - the two cases being in fact the same up to conformal anomalies, which we can safely ignore here since we only study correlation functions of the defect with local operators. The large residual symmetry group preserved by this kind of defects makes them the obvious place to start. Notice however that a systematic study of perturbations around a flat or spherical defect is possible in terms of the displacement operator -- see \emph{e.g.} \cite{Polyakov:2000ti,Billo:2016cpy,Bianchi:2015liz} -- so that complete knowledge of this highly symmetric case is in principle sufficient to study a defect of generic shape. We shall always use the following convention:
\beq
p=\textup{dimension of the defect,}\quad q=\textup{codimension of the defect,}\quad d=\textup{dimension of spacetime,}
\eeq
so that of course $p+q=d.$ The defect symmetry group is then $SO(p+1,1)\times SO(q)$, the first factor accounting for conformal transformations on the defect, the second for rotations around it. The $q=1$ case is degenerate, and we shall sometimes treat it separately.

The ordinary fusion of local operators is hereafter called the bulk OPE, and is schematically denoted as follows:
\beq
\Ocal_1(x_1) \Ocal_2(x_2) \sim \sum_{\Ocal} c_{12\Ocal} \Ocal(x_2)~.
\label{bOPEschem}
\eeq
Local operators of the bulk theory are labeled as usual by their scaling dimension and $SO(d)$ representation:
\beq
\Ocal:\ \{\D, l\}.
\eeq

In the presence of a conformal defect, a new OPE channel opens. A local operator can be fused with the defect. As it can be proven by doing radial quantization centered in a point on the defect, the result of the fusion is a convergent sum over local excitations on the defect, which we call defect operators. We call this the defect OPE channel, and denote it schematically as follows: 
\beq
\Ocal(x) \sim \sum_{\hOcal} b_{\Ocal\hOcal} \hOcal(x^a)~.
\label{dOPEschem}
\eeq
In our conventions, the presence of the defect is understood, defect operators are denoted with a hat, and letters from the beginning of the alphabet ($a,\,b,\dots$) are used to denote directions parallel to the defect. We denote orthogonal directions with letters from the middle of the alphabet ($i,\,j,\dots$).\footnote{This is the notation for a flat defect. For the conventions in the spherical case we refer to subsection \ref{subsec:embedding}.} In the schematic equations \eqref{bOPEschem} and \eqref{dOPEschem}, we kept explicit the bulk-to-bulk and the bulk-to-defect OPE coefficients $(c_{12\Ocal},b_{\Ocal\hOcal})$, but we suppressed all the kinematics and the spin indices. In particular, the defect operators are labeled by their quantum numbers under the defect algebra $so(p+1,1)\times so(q)$: 
\beq
\hOcal:\ \{\hD, \hl, s\},
\eeq
where $\hD$ is the scaling dimension, $\hl$ is the spin under $so(p)$ and $s$ is the spin under $so(q)$. We refer $\hl$ and to $s$ as the parallel and transverse spin respectively.
The identity might appear in the defect OPE, in which case the bulk operator $\Ocal$ acquires an expectation value. Following the literature, we denote this coefficient differently: 
\beq
b_{\Ocal\, 1}\equiv a_{\Ocal}.
\label{onepointb}
\eeq

It is sometimes useful to consider a third OPE channel. Consider a spherical defect. We could replace it by a sum over local operators placed, say, at the center of the sphere \cite{Berenstein:1998ij,Gadde:2016fbj,Fukuda:2017cup}. In correlation functions, this is equivalent by conformal invariance to the fusion of all other operators. In radial quantization, this replacement provides the decomposition of the defect in a complete basis of local operators:
\beq
\ket{\textup{defect}}=\sum_{\Ocal} a_{\Ocal}  \ket{\Ocal}~.
\eeq
The coefficients in the decomposition are the one-point functions in eq. \eqref{onepointb}, up to the kinematics that we are still suppressing. Analogously, a defect excited by a local defect operator generates another state in radial quantization, whose decomposition now involves the $b_{\Ocal\hOcal}$, and so on. Hence, defects are not new states in the Hilbert space of the bulk theory in radial quantization. However, as pointed out, a defect comes equipped with a new Hilbert space generated by defect operators. Given a correlation function involving local operators and defects, we could insert a resolution of the identity in terms of the bulk or the defect Hilbert spaces. The compatibility of the two OPE decompositions is a crossing constraint involving the coefficients $b_{\Ocal\hOcal}$. Let us finally mention that one could also fuse two defects, but we do not treat this problem here: in what follows, the presence of a single conformal defect is always understood.

The focus of this paper is on the simplest of the crossing constraints: the one involving a two-point function of bulk local operators in the presence of a conformal defect. The presence of the defect is denoted with a subscript:
\beq
\braket{\Ocal_1(x_1)\Ocal_2(x_2)}_D~.
\label{2pointschem}
\eeq
The two-point function admits two OPE decompositions: one can plug in the bulk OPE \eqref{bOPEschem} or the defect OPE \eqref{dOPEschem}. The resulting crossing equation can be written in the following way, when $\Ocal_1$ and $\Ocal_2$ are scalar operators:
\beq
\braket{\Ocal_1(x_1)\Ocal_2(x_2)}_D = \sum_{\Ocal} c_{12\Ocal} a_{\Ocal} G_{\Ocal} (x_1,x_2) =  \sum_{\hOcal} b_{\Ocal_1\hOcal} b_{\Ocal_2\hOcal} \hG_{\hOcal}(x_1,x_2)~.
\label{crossschem}
\eeq
The conformal partial waves $G_{\Ocal} (x_1,x_2)$ and $\hG_{\hOcal}(x_1,x_2)$ are fixed by symmetry. In the case of a defect of codimension one ($q=1$), both the bulk and defect channel conformal partial waves are known in closed form when the external primaries are scalars or spin 2 operators \cite{McAvity:1995zd,Rastelli:2017ecj}. In \cite{Billo:2016cpy}, the case of higher codimension was considered, for scalar external operators. In this paper, the defect channel partial waves were found in closed form, while recurrence relations in a lightcone expansion were given for the bulk channel partial waves of symmetric traceless tensors. Furthermore, in the special case $q=2$ the bulk channel partial waves were mapped into those of the ordinary four-point function of local operators.

The two-point function \eqref{2pointschem} has two cross ratios, except in the degenerate case $q=1$. The main purpose of this paper is to define two new pairs of cross ratios, which are convenient in studying the bulk and defect OPE decompositions respectively. The new cross-ratios have properties similar to the radial coordinates defined in \cite{Pappadopulo:2012jk} for the ordinary four-point function, which prompts us to also refer to them as radial coordinates. We define the radial coordinates in section \ref{sec:radial&co}, after a review of the embedding formalism and its application to the study of defect CFTs. Along the way, we present a general classification of the cross ratios involved in a generic correlation function.
Section \ref{sec:scalar} is dedicated to the conformal partial waves for scalar external operators. We show how to compute the partial waves in the radial expansion by applying methods which are routinely used in the case of the four-point function: a recursive solution to the Casimir equation \cite{Hogervorst:2013sma}, and the Zamolodchikov expansion \cite{Zamolodchikov:1985ie}.
In section \ref{sec:convergence} we discuss the convergence properties of the bulk and the defect OPEs. We show that the radius of convergence of the radial expansion equals the region of convergence of the OPE. In the case of identical external operators, we also give an estimate of the rate of convergence of the defect OPE decomposition, following in the footsteps of \cite{Pappadopulo:2012jk}. Finally, in subsection \ref{subsec:landau}, we contrast the radial expansion with the expansion of the bulk channel blocks in the cross ratio $\xi$ -- see eq. \eqref{crossxizeta} below -- which is normally used in the defect CFT literature. It is remarkable that the latter stops converging at $\xi=1$, which is precisely the point used so far in the numerical bootstrap \cite{arXiv:1210.4258,arXiv:1502.07217} with either the extremal functional \cite{arXiv:0807.0004} or the determinant method \cite{Gliozzi:2013ysa}. This is not an issue for the bootstrap of codimension one defects, since the corresponding blocks are known in closed form. The radial coordinates are instead crucial to bootstrap more general defects.
A certain amount of technical details and explicit results is relegated to the appendices.
Finally a  Mathematica notebook that computes the bulk channel conformal blocks is included in the submission of the paper.

\section{Defect CFTs and radial coordinates}
\label{sec:radial&co}
\subsection{Embedding space formalism}
\label{subsec:embedding}

We begin this section by reviewing the embedding space formalism for $d$ dimensional CFTs \cite{SpinningCC}. We uplift each point in the physical space $\mathbb{R}^d$ to a point on the null cone of the embedding space $\mathbb{R}^{1,d+1}$ defined by
\be
\mbox{null cone}=\left\{ P^M\equiv\left( P^{0},P^\mu,P^{d+1}\right)\in \mathbb{R}^{1,d+1}\ 
:
 \
P^M P_M=0 \right\} 
\ ,
\ee
where $P^M P_M \equiv - (P^{0})^2+P^{\m} P^{\n} \d_{\m \n}+(P^{d+1})^2$. 
The physical space is then described by the set of rays $P \sim \a P$ (with $\a>0$). In particular, $x \in \mathbb{R}^d$ is obtained by projecting $P$ onto the Poincar\'e section $P^{d+1}+P^{0}=1$, parametrized by
\be
\label{PoincareSec}
P_{\mbox{\scriptsize Poincar\'e}}=\left(\frac{1+x^2}{2} , x^\m , \frac{1-x^2}{2} \right) \ .
\ee
Other conformally flat spaces are recovered by projecting onto different sections.
For instance, consider the section $\delta_{\mu \nu}P^\mu P^\nu=1$. This is naturally parametrized by a Euclidean time $\tau \in \mathbb{R}$ and a unit vector $n^\mu \in S^{d-1} \subset \mathbb{R}^d$,
\be
\label{cyl_section}
P_{\mbox{\scriptsize cyl}}=(\cosh \tau,n^\mu,-\sinh \tau)\ .
\ee
The induced line element is 
$
ds^2=d\tau^2+d\Omega^2_{S^{d-1}}
$, which is the metric on the cylinder  $\mathbb{R}\times S^{d-1}$. The Poincaré and the cylinder sections are simply related:
\beq
P_\textup{Poincaré} = e^\tau P_\textup{cyl},\qquad r=e^\tau.
\label{PoinToCyl}
\eeq
The usefulness of the embedding space formalism stems from the following fact: conformal transformations of the physical space act as Lorentz transformations in embedding space.

\begin{figure}
\centering
\qquad\qquad\qquad\qquad
\graphicspath{{Fig/}}
\def\svgwidth{7 cm}
\subfloat{\ \ \   }
\subfloat{\input{Embedding1.pdf_tex}}
\subfloat{\ \ \ \ \ \  }
\subfloat{\input{Embedding2.pdf_tex}}
\caption{
Both pictures represent the example of a $(d+2)=3$ dimensional embedding space, labelled by $P^M=(P^0,P^1,P^{2})$. 
We show the null cone, the defect plane and  the Poincar\'e section (in red).
In the left picture, the $P^-$ axis lies on the defect plane. The latter intersects the Poincar\'e  section in a single point (in yellow), which in physical space corresponds to a flat defect.
In the right picture, the $P^1$ direction is parallel to the defect plane, while $P^-$ does not lie on it. The defect plane intersects the Poincar\'e section along $P^1$ in two points. This is a spherical defect centred in the origin.
}
\label{fig:3dcones}
\end{figure}
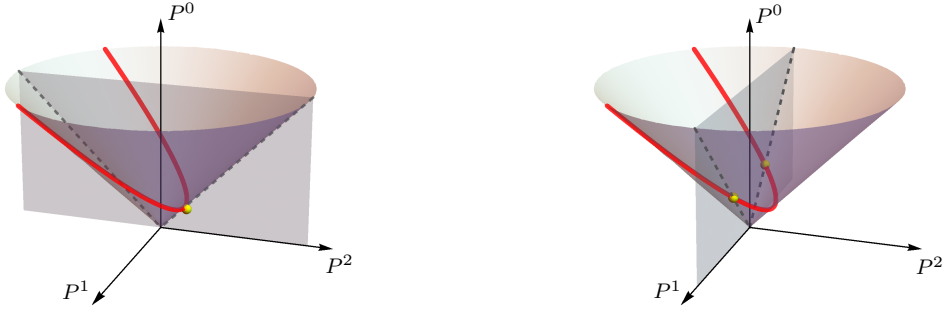

It is not hard to understand how to describe in this formalism a defect of dimension $p$ and codimension $q$, embedded in a $p+q=d$ dimensional CFT \cite{Billo:2016cpy}. In the  embedding space $\mathbb{R}^{1,d+1}$, the defect is a $p+2$ dimensional time-like plane which goes through the origin and intersects the $(d+1)$-dimensional light-cone. Indeed, the plane preserves the defect conformal group $SO(p+1,1)\times SO(q)$.  The projection of the defect plane onto the Poincaré section is generically a $p$-sphere. A flat defect is obtained as a special case, when the axis $P^-=P^{0}-P^{d+1}$ is contained in the defect plane.\footnote{A proof of this statement goes as follows. A plane in embedding space is described by $q$ linear equations $c^i_M P^M=0$, $i=1,\dots,q$. Projected onto the Poincaré section, the system defines the defect in Euclidean $d-$dimensional space. For it to be flat, the equations should still be linear in $x^\mu$, so the term $c_-^i P^-=c^i_- x^2$ must vanish.} Since all spheres are conformally equivalent to each other, one can keep in mind a single configuration, without loss of generality.\footnote{For instance, the action of $P_\mu=J_{\mu-}$ leaves the $P^-$ axis invariant and so translates a flat defect without deforming it, etc. A detailed analysis can be found, for instance, in \cite{Gadde:2016fbj}.} In practice, it will be useful to consider explicitly two situations: a flat defect passing through the origin, and a spherical defect centered in the origin, whose radius we set to one. It is easy to see that the former is realized by choosing both $P^{0}$ and $P^{d+1}$ among the parallel directions, while the latter corresponds to a defect plane lying at $P^{d+1}=0$. The two situations are represented in fig. \ref{fig:3dcones}.

It is convenient to introduce projectors $\Pp $ and $\Pt$ onto the space parallel and orthogonal to the defect plane.  
Correspondingly, in addition to the full $d+2$-dimensional  scalar product 
\be
P\cdot Q \equiv \sum_{M} P^M Q_M \ ,
\ee
we introduce the scalar products in parallel and transverse directions
\ba
\label{Pdot}
P{\pdot} Q & \displaystyle\equiv P \cdot \Pp \cdot Q 
& \qquad \mbox{(parallel)}\ , \\
\label{Tdot}
P {\tdot } Q  &\displaystyle \equiv P \cdot \Pt \cdot Q 
\ , &\qquad \mbox{(orthogonal)}\ .
\ea
We can select the shape of the defect in physical space by specifying the form of the projectors  $\Pp$ and $\Pt$:
\be
\label{ProjectorsEmbedding}
 \begin{array}{l l}
\Pt=\textup{diag}(0,\underbrace{0 \dots 0}_{p},\underbrace{1 \dots 1}_{q},0) &\ \Longrightarrow \  \mbox{flat defect} \ , \\
\Pt=\textup{diag}(0,\underbrace{0 \dots 0}_{p+1},\underbrace{1 \dots 1}_{q-1},1)& \ \Longrightarrow \   \mbox{spherical defect} \ .
\end{array} 
\ee
In other words, $P^{d+1}$ is among the orthogonal coordinates for a flat defect and the parallel ones for a spherical defect.
Similarly, $\Pp$ follows from the relation $\Pp+\Pt=\textup{diag}(-1,1,\dots,1)$.
The intersection of the defect plane with a different section of the cone yields the image of a spherical or flat defect under Weyl transformations. 
In practice to project onto any section -- \emph{e.g.} the Poincar\'e or the cylindrical section -- we just need to choose a parametrization of the points -- \emph{e.g.} (\ref{PoincareSec}) or (\ref{cyl_section}) -- and to specify a form for the projector $\Pi_{\,\wbullet}$ according to (\ref{ProjectorsEmbedding}).

These definitions are enough to study any defect conformal field theory. However, it is useful to introduce some extra notations which simplifies the expressions in real space. We define a splitting of the real space indices $\m=1, \dots ,d$ of a vector $x \in \mathbb{R}^d$ into parallel and transverse directions labelled respectively by the letters $a,b,\dots$ and $i,j,\dots$. 
We consequently define three different scalar products\footnote{With some abuse of notation, the products in embedding space and in physical space will be denoted by the same symbols. We hope that this is not a source of confusion.
} for vectors $x,y \in \mathbb{R}^d$: the full product $x \cdot y \equiv x^\m y^\n \d_{\m \n}$ and the parallel and transverse products 
\ba
\label{pdot}
x{\pdot} y & \displaystyle\equiv x \cdot \pp \cdot y \equiv \sum_{a} x_a \,  y_a  \ , & \qquad \mbox{(parallel)}\ , \\
\label{tdot}
x {\tdot } y  &\displaystyle \equiv x \cdot \pt \cdot y \equiv \sum_{i} x_i \, y_i \ , &\qquad \mbox{(orthogonal)}\ .
\ea
Here we denoted by $\pi_{\pdot}$ and $\pi_{\tdot}$ the projectors onto parallel and orthogonal components. The products satisfy the relation $x \cdot y= x  \pdot y+x  \tdot y$.
Again, the splitting between parallel and transverse directions is different if we consider flat or spherical defects,
\be
\label{ProjectorsRealSpace}
 \begin{array}{lcl l}
\pt=&\pi_{q}&\equiv \textup{diag}(\underbrace{0 \dots 0}_{p},\underbrace{1 \dots 1}_{q}) &\ \Longrightarrow \  \mbox{flat defect} \ , \\
\pt=&\pi_{q-1}&\equiv \textup{diag}(\underbrace{0 \dots 0}_{p+1},\underbrace{1 \dots 1}_{q-1})& \ \Longrightarrow \   \mbox{spherical defect} \ .
\end{array} 
\ee
Notice that in the spherical case we call `parallel' the $p+1$ directions in which the defect is embedded. These directions are in fact parallel to the defect plane in embedding space. 
Of course, the definition of $\pp$ follows from $(\pp+\pt)_{\m \n}=\d_{\m \n}$.
To avoid confusion, we will often remind the reader which convention for the scalar product we are using.

\subsection{Correlation functions and cross-ratios}
\label{Sec:Cross_Ratios}

The aim of this paper is to define a convenient set of cross ratios for the study of a two-point function of bulk primaries. However, for completeness, we begin by a general classification. Consider a correlation function of $n$ bulk primaries $\Ocal_i$ with dimensions $\D_i$ and $m$ defect primaries $\hat \Ocal_i$ with dimensions $\hD_i$. For notational simplicity we refer to scalar operators:
\be
\label{Generic_Correlation_Function}
\left \langle 
 \prod_{i=1}^n
\Ocal_i(P_i)
 \prod_{j=n+1}^{n+m}
\hat \Ocal_{j}(\hP_{j}) 
\right \rangle \ .
\ee
We denoted as $\hP_j$ the insertion points of the defect operators: it is understood that $\Pt \cdot \hP =0$.
We would like to classify the conformal invariant cross ratios for the correlation function \eqref{Generic_Correlation_Function}. Let us start by counting them. One way is to organize the coordinates $P^M$ of the $n$ bulk points and the $m$ defect points in a rectangular matrix $(d+2)\times (n+m)$. One can then reduce the number of non vanishing components by applying elements of $SO(p+1,1)\times SO(q)$. Of course, the number of independent components is also constrained by projectiveness ($P^M~\sim~\lambda P^M$) and nullity ($P^2=0$). The final count depends on the number columns and of rows in the orthogonal and parallel subspaces. We summarize it in table \ref{cross_ratios}.
\begin{table}[h]
\centering
\begin{tabular}{|c | c | c |}
  \hline 
 \hspace{0.5 cm}
 $ n+m $ 
  \hspace{0.5cm} 
 & \hspace{0.5 cm} $n \hspace{0.5cm}$
 &
   \phantom{ \LARGE |}
  \hspace{0.5 cm} 
  Number of cross ratios
    \hspace{0.5 cm}
      \phantom{ \LARGE |}
      \\  
      \hline
 \multirow{3}{*}{$\geq p+2$  } 
 & 
 $\geq q$
  & 
  \phantom{ \LARGE $\big|$}
 $
\frac{ 2d n+2p m -q(q-1)-(p+2)(p+1)}{2} $
  \phantom{ \LARGE $\big|$}
  \\
 &  
$ < q$
 &
  \phantom{ \LARGE $\big|$}
 $
\frac{n(n+1)+2p(n+m)-(p+2)(p+1)}{2}$ 
  \phantom{ \LARGE $\big|$}
  \\
 \hline
 \multirow{ 3}{*}{$  < p+2$}   
 & 
  $\geq q$
  & 
    \phantom{ \LARGE $\big|$}
  $
  \frac{2 q n-q(q-1)+(n+m)(n+m-3)}{2}
  $  
    \phantom{ \LARGE $\big|$}
  \\      
  & 
  $ < q$
   & 
  \phantom{ \LARGE $\big|$}
   $
   \frac{n(n+1)+(n+m)(n+m-3)}{2}$
  \phantom{ \LARGE $\big|$}
   \\
 \hline
\end{tabular}
\caption{Number of cross ratios in \eqref{Generic_Correlation_Function}: $n$ and $m$ are the numbers of bulk and defect operators respectively. Notice that the middle cases involve restrictions on the relation between dimension and codimension: $p+2 \lessgtr q+m$ respectively.}
\label{cross_ratios}
\end{table}

We now explicitly present the set of cross ratios, providing $p$ and $q$ are large enough, that is, corresponding to the last row of table \ref{cross_ratios}. It is convenient to treat separately the case $n=0$. If all the points are on the defect, one can simply use the usual basis of cross ratios appropriate for a $p$-dimensional CFT: 
\be
u_{ i  j  k  l} \equiv  \frac{ (\hP_i \pdot \hP_j)(\hP_k \pdot \hP_l) }{(\hP_i \pdot \hP_k)(\hP_j \pdot \hP_l) } \ .
\ee
When at least a bulk point is involved, a basis can be chosen as follows. We define three classes of cross ratios, depending on the number of defect points involved, zero, one, or two:
\begin{subequations}
\begin{align}
u^{\star}_{ij} &\equiv \frac{ (P_i \star  P_j) }{\sqrt{(P_i \pdot P_i)(P_j \pdot P_j) } }\ , \qquad \star=\pdot,\tdot\ , \\
u_{\hi,jk} &\equiv 
\frac{  (\hat{P}_i \pdot  P_j)  }{(\hat{P}_i \pdot P_k)  } 
\sqrt{\frac{(P_k \pdot P_k)}{(P_j \pdot P_j) }} 
\ , \\
u_{ \hi \hj , k} &\equiv  \frac{ (\hP_i \pdot \hP_j)(P_k \pdot P_k) }{(\hP_i \pdot P_k)(\hP_j \pdot P_k) } \ .
\end{align}
\label{eq:cross-ratios}
\end{subequations}
The minimum number of bulk operators associated to each class is consistent with the fact that one-point functions, bulk-to-defect and defect-to-defect two-point functions are all fixed by conformal invariance. The cross ratios \eqref{eq:cross-ratios} are not all independent. In particular, we signal the relations $u_{\hi, k\,  k+l}=\prod_{j=k}^{k+l-1} u_{\hi,j \, j+1}$ and $ u_{\hi \hj, k}=  u_{\hi \hj ,l} u_{\hi,kl}u_{\hj,kl}$. 
A set of cross ratios which are independent when $p$ and $q$ are large enough can be chosen as follows:
\be
\label{All_Cross_Ratios}
\begin{array}{| l l | c |}
\hline
&
$ \phantom{ \LARGE |}$ 
\qquad
\mbox{Cross Ratios}
&
$ \phantom{ \LARGE |}$ 
\mbox{Number}
$ \phantom{ \LARGE |}$
\\
\hline
$ \phantom{ \LARGE |}$
u^{\star}_{ij}  
\quad
 &\star=\pdot,\tdot
  \ \ \ \ 1\leq i<j\leq n
 &
 $ \phantom{ \LARGE |}$
 n(n-1)
 $ \phantom{ \LARGE |}$
\\
$ \phantom{ \LARGE |}$
u_{\hi,j\,  j+1} 
\quad
 & i=n+1,\dots, n+m
 \ \ \ \ j=1,\dots n-1
\qquad 
&
$ \phantom{ \LARGE |}$
 m (n-1)
 $ \phantom{ \LARGE |}$
\\
$ \phantom{ \LARGE |}$
u_{\hi \hj , 1}   &n+1\leq i<j\leq n+m 
\qquad 
&
$ \phantom{ \LARGE |}$
 \frac{m(m-1)}{2}
 $ \phantom{ \LARGE |}$
 \\
 \hline
\end{array}
\ee
Notice that by summing the numbers in the last column of \eqref{All_Cross_Ratios} we recover the counting in the last row of table \ref{cross_ratios}.

\subsection{The $\rho$ coordinates}
\label{subsec:rho}

The focus of this paper is the two-point function of bulk primaries:
\be\label{newStructEmi}
\langle \Ocal_1(P_1) \Ocal_2(P_2) \rangle_D = \frac{1}{(P_1 \tdot P_1)^{\frac{\D_1}{2}}(P_2 \tdot P_2)^{\frac{\D_2}{2}}} 
 f(u_1, u_2) \ .
\ee
This correlator is a function of two cross ratios, generically denoted as $u_1,\, u_2$.  As we recalled in the introduction - see eq. \eqref{crossschem} - a two-point function admits two partial wave decompositions.
We shall now define two pairs of cross ratios, which are useful in studying the defect OPE and the bulk OPE respectively. As we shall see in the following, the cross ratios have analogous properties to the $\rho$ coordinates defined in \cite{Pappadopulo:2012jk} and studied in detail in \cite{Hogervorst:2013sma}.
\\
The new cross ratios are most naturally defined on the cylinder $S^{d-1}\times \mathbb{R}$. Choosing how to embed the defect on the cylinder is equivalent to the choice of the origin in radial quantization on the plane $\mathbb{R}^d$. In an ordinary CFT, translational invariance implies that the spectrum on the sphere does not depend on this choice. On the contrary, when a defect is present, radial quantization around a point on the defect yields the spectrum of defect operators. Now, we first define coordinates suitable for the study of the bulk OPE, and then we turn to the defect OPE.

\begin{figure}
\centering
\includegraphics[scale=0.7]{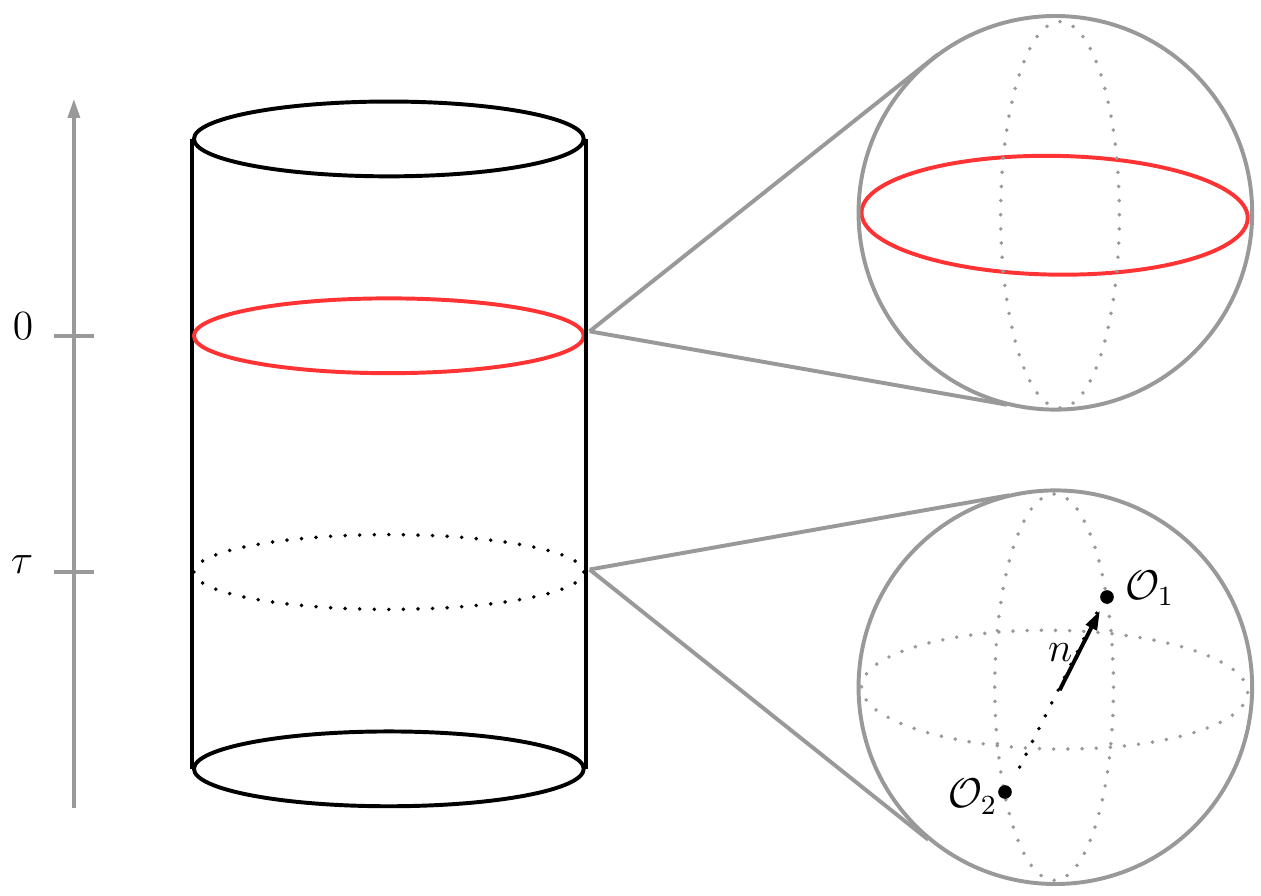}
\caption{The configuration corresponding to eq. \eqref{PiOnTheCylinder1}. The defect is marked in red, and lies on one equator of the sphere at $\tau=0$. The local operators are placed at equal generic time, on the orthogonal equator, in opposite points.}
\label{fig:cylinder_B}
\end{figure}

\begin{figure}[h]
\centering
\includegraphics[scale=0.78]{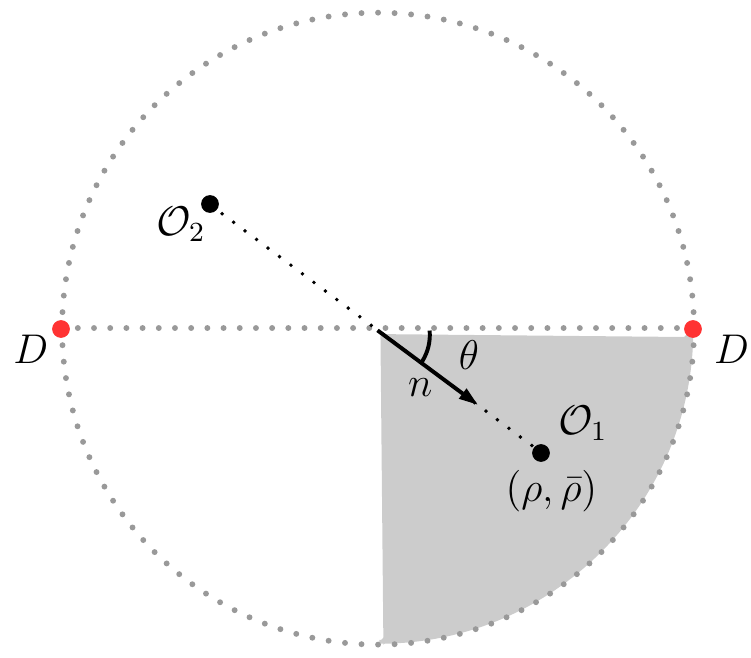}
\caption{The configuration corresponding to eq. \eqref{B_Poin_Conf}. The defect is spherical and orthogonal to the plane drawn in the figure, and crosses it at the position marked by the red dots. The operators $\mathcal{O}_1$ and $\mathcal{O}_2$ sit at the same radius $r$, and the position of $\mathcal{O}_1$ is parametrized by the complex coordinates $(\rho,\bar{\r})$. The fundamental domain $\mathcal{D}$ is highlighted in gray.}
\label{fig:rho}
\end{figure}

\paragraph{Bulk channel}
If we are interested in the bulk channel conformal block decomposition, it is convenient to choose the vacuum of the homogeneous CFT both as in and as out-state. In embedding space, we choose the cylindrical section \eqref{cyl_section} and the splitting suitable for a spherical defect in \eqref{ProjectorsEmbedding}, so that on the cylinder the defect is an $S^p$ sitting at $\tau=0$, see fig. \ref{fig:cylinder_B}.
We further pick the position of the two primaries as follows: 
\begin{align}
\begin{split}
\label{PiOnTheCylinder1}
P_1=(\cosh \t, n,-\sinh \t)\ , \qquad
P_2=(\cosh \t,  -n,- \sinh \t)\ ,
\end{split}
\end{align}
 where $n$ is a unit vectors in $\mathbb{R}^d$. 
 This naturally leads to the definition of the following cross ratios:
\be
 r \equiv  e^{\t} \ , \qquad \eta^2
 \equiv n \pdot n 
 \ ,
 \label{bulkreta}
\ee
where the parallel product is the one in physical space, eq. \eqref{pdot}, and is defined according to the spherical splitting in \eqref{ProjectorsRealSpace}.
It is sometimes best to think in terms of the complex version of the coordinates:
\beq
\rho=r e^{\ii \theta},\qquad \bar{\rho}=r e^{-\ii \theta},
\qquad \eta=\cos\theta.
\label{bulkrho}
\eeq
When we project the configuration \eqref{PiOnTheCylinder1} onto the Poincaré section, that is we simply use eq. \eqref{PoinToCyl}, we obtain the configuration shown in fig. \ref{fig:rho}, with
\be
\label{B_Poin_Conf}
P_1=\left(\frac{1+r^2}{2}, r n, \frac{1-r^2}{2}\right) \, 
\qquad
P_2=\left(\frac{1+r^2}{2},-r n,\frac{1-r^2}{2}\right).
\ee
The configuration is analogous to the one of the $\rho$ coordinate for the four-point function, but the fundamental domain is different: $(r,\eta)$ can be restricted to lie in the region
\beq
\mathcal{D}=\{|\rho|\leq1,\Re\rho\geq0,\Im\rho\leq0\},
\label{regD} 
\eeq 
because of the symmetries $r\to1/r$ and $\eta\to-\eta$. The second symmetry is absent in the case of a four-point function of non-identical operators: it is implemented, for instance, by a rotation in the plane formed by the axis $\Re\rho$ and one of the other axes that intersect the defect. This is an element of $SO(p+1,1)$. The constraint on the sign of $\Im\rho$ follows from the fact that $\theta$ and $-\theta$ map to the same value of $\eta$.\footnote{In fact, one can map $\theta\to-\theta$ via a rotation involving two orthogonal directions. This transformation reduces to parity when $q=2$. If parity is not a symmetry the correlator with a defect of codimension 2 may depend separately on $e^{\ii\theta}$ and $e^{-\ii\theta}$.}

\begin{figure}[h]
\centering
\includegraphics[scale=0.7]{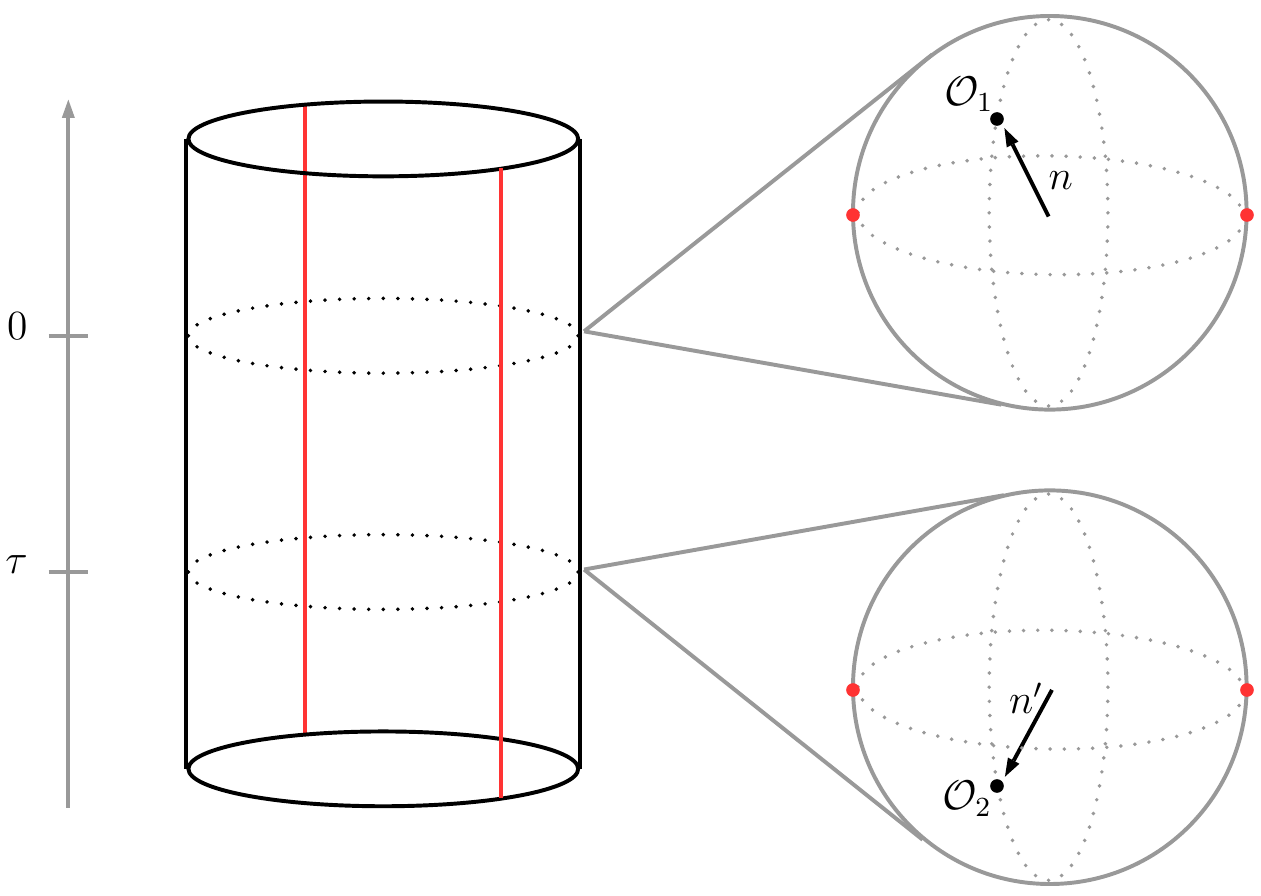}
\caption{The configuration corresponding to eq. \eqref{D_Cyl_Conf}. The red lines on the cylinder mark the position of the defect. Constant time slices are spheres, which in this three dimensional example are marked by the defect in two opposite poles. The operators $\mathcal{O}_1$ and $\mathcal{O}_2$ live on the equator of two spheres inserted at time $0$ and $\tau$ respectively.}
\label{fig:cylinder_D}
\end{figure}

\begin{figure}
\centering
\includegraphics[scale=0.78]{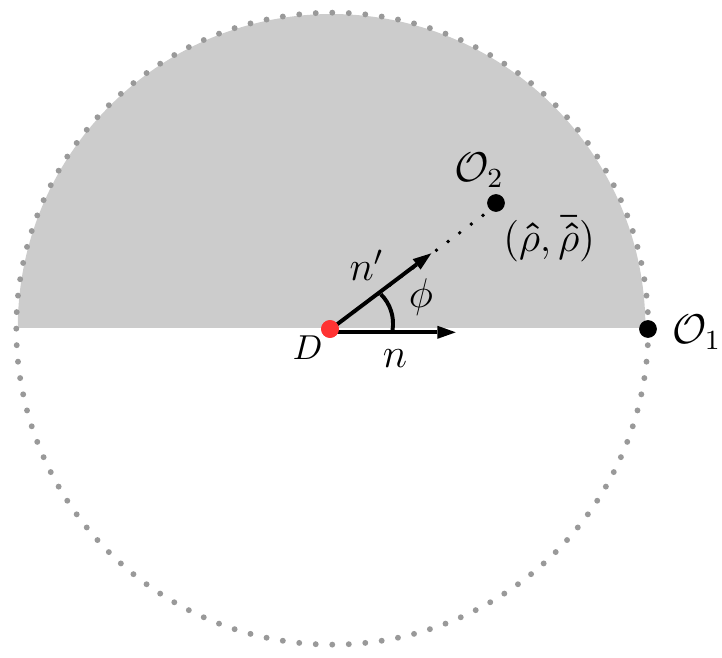}
\caption{
The configuration corresponding to eq. \eqref{D_Poin_Conf}. The defect is flat and orthogonal to the plane drawn in the figure, and crosses it at the position marked by the red dot. The operator $\mathcal{O}_1$ sits at unit radius, while the position of $\mathcal{O}_2$ is parametrized by the complex coordinates $(\hrho,\hrhobar)$. The fundamental domain $\hat{\mathcal{D}}$ is highlighted in gray.}
\label{fig:rhohat}
\end{figure}

\paragraph{Defect channel}
Turning our attention to the defect conformal block decomposition, the natural choice is to center the radial quantization on a point belonging to a flat defect. The Hilbert space of the theory is defined on an $S^{d-1}$ marked by the defect along an $S^{p-1}$. The in and out-states are picked to be the ground state in this Hilbert space, which we refer to as the defect vacuum.\footnote{The latter requirement is in fact the only crucial one: the same result is obtained by doing North-South pole quantization \cite{Rychkov:2016iqz}, and choosing two points on the defect as the North and South poles.} On the cylinder, the defect is mapped to a lower dimensional cylinder $S^{p-1}\times \mathbb{R}$, see fig. \ref{fig:cylinder_D}. 
In embedding space, we now choose the splitting suitable for a flat defect in \eqref{ProjectorsEmbedding}, which suggests to split the coordinates of the unit vector $n^\mu=(n^a,n^i)$ according to the first of the \eqref{ProjectorsRealSpace}  as well. We choose to place the operator 1 in $(\tau=0,n^a=0)$ and the operator 2 at a generic value of $\tau$, but still at $n^a=0$. In embedding space,
\be
\label{D_Cyl_Conf}
P_1=(\cosh 0, \underbrace{0 \dots 0}_{p}, n,-\sinh 0) \, 
\qquad
P_2=(\cosh \t, \underbrace{0 \dots 0}_{p}, n',-\sinh \t)
\ee
where $n$ and $n'$ are unit vectors in the sphere $S^q$.
We take as cross ratios the coordinates of $P_2$, as follows:
\be
\hr\equiv e^{\t} \, , \qquad \heta\equiv  n \tdot  n' 
\ ,
\label{defectreta}
\ee
where we used the the transverse dot eq. \eqref{tdot}, still adapted to the flat defect.
Also in this case, let us define the $\hrho$-coordinates as
\beq
\hrho=\hr e^{\ii \phi}, \qquad \hrhobar=\hr e^{-\ii \phi},\quad \heta=\cos\phi.
\label{defectrho}
\eeq
If we project eq. \eqref{D_Cyl_Conf} to the plane, we obtain
\be
\label{D_Poin_Conf}
P_1=(1, \underbrace{0 \dots 0}_{p}, n, 0) \, 
\qquad
P_2=\left(\frac{1+\hr^2}{2}, \underbrace{0 \dots 0}_{p}, \hr n',\frac{1-\hr^2}{2}\right)~.
\ee
We depict the corresponding configuration in fig. \ref{fig:rhohat}.
Despite the similarity with the $(z,\bar{z})$ parametrization of the 4 point function without defect, the range of the $\hrho$ coordinate is restricted to the region
\beq
\hat{\mathcal{D}}=\{|\hrho|\leq1,\Im\hrho\geq 0\}.
\label{regDhat}
\eeq 
Indeed, an inversion $\tau\to-\tau$ maps every point with $\hr>1$ to this fundamental region,\footnote{In fact, a rotation by an angle $\pi$ in the $(P_{0},P_A)$ plane, $A$ being any parallel direction, has the same effect. Parity invariance is therefore irrelevant for this statement.} and a rotation in transverse space exchanges $\hrho$ and $\hrhobar$.\footnote{The latter is reduced to parity in codimension 2. If parity is broken, correlation functions might be functions of $e^{\ii \phi}$ and $e^{-\ii \phi}$ separately.} This is to be contrasted with the four-point function. In the latter case, the inversion has already been used to send an insertion to infinity, and is unavailable.

We would further like to point out that the relation $\r(\hrho)$ takes the following familiar form:
\beq
\r=\frac{1-\sqrt{\hrho}}{1+\sqrt{\hrho}},\qquad 
\hrho=\frac{(1-\r)^2}{(1+\r)^2}~.
\label{rho(hrho)}
\eeq
Analogous formulae hold for $\bar{\r}(\hrhobar)$.
This is the same as the usual $\rho(z)$ relation four the four-point function, once we replace $z \to 1-\hrho$. Notice that, while this relation is invertible everywhere in the complex $\hrho$ plane, the disc $|\hrho|<1$ is mapped to the half-disc with $\Re \rho>0$, concordantly with the fundamental domain discussed for the pair $(r,\eta)$. Furthermore, $\Im \hrho>0$ is mapped into $\Im \rho<0$, which is why we chose $\Im \rho<0$ as the fundamental domain for $\rho$ in eq. \eqref{regD}. This is done for consistency, but of course it has very little importance in the following.\footnote{An alternative option is to change the map \eqref{rho(hrho)} so that it interchanges holomorphic and antiholomorphic coordinates.}
For reference, let us also report the relation between $(r,\eta)$ and $(\hr,\heta)$:
\begin{align}
r&= \sqrt{\frac{1+\hr-\sqrt{2\hr (1+\heta)}}{1+\hr+\sqrt{2\hr (1+\heta)}}}~,&
\eta&=\frac{1-\hr}{\sqrt{1+\hr^2-2\hr \heta}}~,
\label{r(rhat)}  \\
\hr &=\frac{1+r^2-2r\eta}{1+r^2+2r\eta}~,&
\heta &= \frac{1-6r^2+4 \eta ^2 r^2+r^4}{1+2 r^2-4 \eta ^2 r^2+r^4}~.
\label{rhat(r)}
\end{align}

Let us finally mention the degenerate case of a codimension one defect. Only one cross ratio survives in this case. Indeed, the transverse and parallel products in physical space - eqs. \eqref{pdot}, \eqref{tdot} - become trivial: for a flat defect $x\tdot y= x^dy^d$, while for a spherical defect $x\pdot y=x\cdot y$. It follows that $\eta=1$ and $\heta=\pm1$ identically. If the defect is a boundary, $\heta=1$, while $\heta=-1$ for a correlation function of operators placed on opposite sides of an interface. It is worth mentioning that, via the so-called folding trick, a pair $(\textup{CFT}_1,\textup{CFT}_2)$ glued at an interface is equivalent to a boundary conformal field theory. The trick consists in applying a reflection only to, say, $\textup{CFT}_2$. The kinematic constraints on correlation functions are then the same as those for the product theory $\textup{CFT}_1\times \overline{\textup{CFT}}_2$ with the interface now replaced by a boundary.

\subsubsection{Relation with other cross ratios}
In \cite{Billo:2016cpy}, the following pairs of cross ratios were used:\footnote{In fact, in \cite{Billo:2016cpy} $\chi$ was the inverse of the one defined here, up to a factor two: $\chi_\textup{here}=2/\chi_\textup{there}$. We redefined it here so that $\chi\in [0,1]$ in Euclidean signature. Likewise, we set $\zeta_\textup{here}=2\zeta_\textup{there}$, so that $\zeta\in [0,1]$ as well. Finally, $\xi_\textup{here}=\xi_\textup{there}/4$. This lasr redefinition is in agreement with the conventions used in older literature \emph{e.g.} \cite{McAvity:1995zd}.}
\beq
\chi=-\frac{1}{u_{12}^\pdot}=-\frac{(P_1\wbullet P_1)^{\frac{1}{2}}(P_2\wbullet P_2)^{\frac{1}{2}}}{P_1\gbullet P_2}~,\qquad
\cos\phi= u_{12}^\tdot=\frac{P_1\wbullet P_2}{(P_1\wbullet P_1)^{\frac{1}{2}}(P_2\wbullet P_2)^{\frac{1}{2}}}~,
\label{crosschiphi}
\eeq
and
\beq
\xi=-\frac{1}{2}(u_{12}^\pdot+u_{12}^\tdot)=-\frac{P_1\cdot P_2}{2(P_1\wbullet P_1)^{\frac{1}{2}}(P_2\wbullet P_2)^{\frac{1}{2}}}~,\qquad \zeta = \, \frac{1-\cos \phi}{2\,\xi}~.
\label{crossxizeta}
\eeq
The two pairs are related by $\chi=1/(2\,\xi+\cos\phi)$. The pair of coordinates $(\chi,\cos\phi)$ is well suited to study the defect OPE, since the defect OPE limit corresponds to $\chi\to0$ at fixed $\cos\phi$. On the other hand, the two bulk operators collide when $\xi\to0$ at fixed $\zeta$, which makes the $(\xi,\zeta)$ pair more useful when dealing with the bulk OPE. Being simple rational functions of invariants in embedding space, $\chi,\,\cos\phi,\,\xi$ and $\zeta$ are especially useful in explicit, say perturbative, computations, since correlation functions are easily recast as functions of a subset of these cross ratios. For reference purposes, and in order to make some comments, let us write explicitly the changes of coordinates between the cross ratios in eq. \eqref{crosschiphi}, \eqref{crossxizeta} and $\rho$, $\hrho$.

We begin with $\rho$,
\beq
\chi = \frac{1+r^4+2\left(1-2 \eta ^2\right) r^2}{1+r^4+2\left(1+2 \eta
   ^2\right) r^2}~, \qquad
\cos\phi= \frac{1-6r^2+4 \eta ^2 r^2+r^4}{1+2 r^2-4 \eta ^2 r^2+r^4}~,
\label{chitor}
\eeq
\beq
\xi=\frac{4 |\r|^2}{|1-\r|^2|1+\r|^2}= \frac{4 r^2}{1+2 r^2-4 \eta ^2 r^2+r^4}~, \qquad
\zeta = 1-\eta^2~.
\label{xitor}
\eeq
and we continue with $\hrho$:
\beq
\chi = \frac{2\hr}{1+\hr^2}~, \qquad
\cos\phi= \heta~,
\label{chitohr}
\eeq
\beq
\xi= \frac{|1-\hrho|^2}{4|\hrho|}=\frac{1}{4}\left(\hr-2 \heta +\frac{1}{\hr}\right)~, \qquad
\zeta = \frac{2\hr (1-\heta)}{1-2\hr \heta+\hr^2},
\label{xitohr}
\eeq

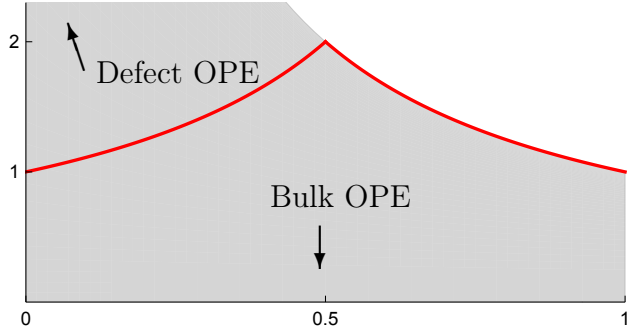
\begin{figure}[t]
\centering
\begin{overpic}[width=0.5\textwidth]{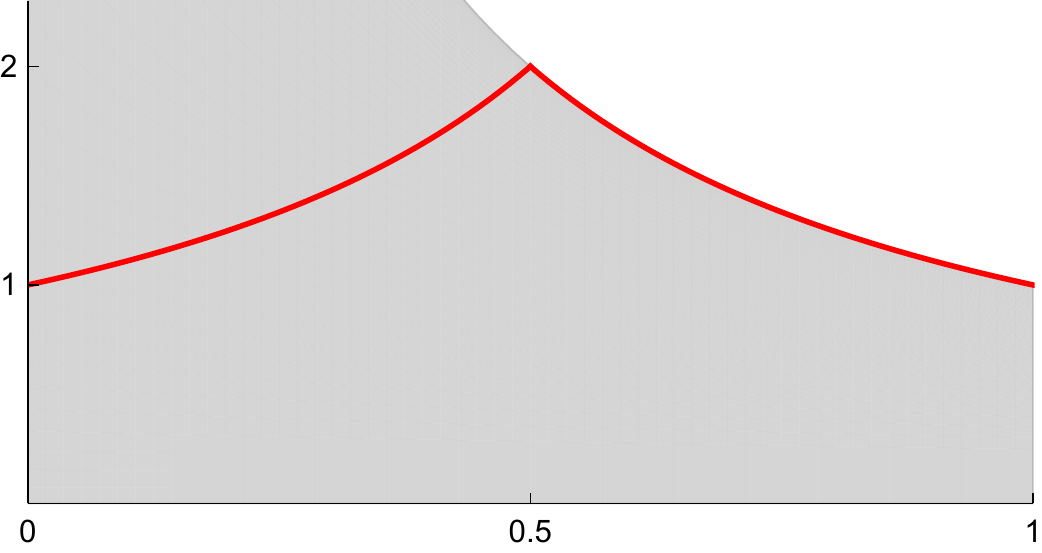}
\put(14,40){Defect OPE}
\put(42,20){Bulk OPE}
\thicklines
\put(12,42){\vector(-1,3){2.5}}
\put(50,17){\vector(0,-1){7}}
\end{overpic}
\caption{The physical region in $(\xi, \zeta)$ coordinates (in gray) is bounded above by the curve $\xi=1/\zeta$, corresponding to $\cos\phi=-1$. The red curve denotes the radius of convergence of the expansion of the bulk channel conformal blocks in powers of $\xi$: the left branch has equation $\xi=1/(1-\zeta)$ - see subsection \ref{subsec:landau}. 
\label{fig:xizetadomain}
}
\end{figure}

Let us make a few comments. First of all, the relations \eqref{chitohr} are invertible in the region $\hat{\mathcal{D}}$ in eq. \eqref{regDhat}, and the relations \eqref{xitor} in the region $\mathcal{D}$ in eq. \eqref{regD}. This is in accordance with the discussion in the previous section. Secondly, one can extract the physical range of each independent pair of cross ratios directly from the definitions \eqref{crosschiphi}, \eqref{crossxizeta}, or from the maps to the $\rho$, $\hrho$ coordinates. In particular, $\chi$ and $\cos\phi$ vary independently: the Euclidean domain is the rectangle $(\chi,\cos\phi)\in [0,1]\times [-1,1]$. The domain of the $(\xi,\zeta)$ pair is slightly more complicated, and so is the one of $(\xi,\cos\phi)$. We draw the former in fig. \ref{fig:xizetadomain}.
This is one reason to prefer the $(r,\eta)$ pair to study the bulk OPE. As a last remark, our definition of $(r,\eta)$, eq. \eqref{bulkreta} is mainly motivated by geometric considerations. However, formulae \eqref{chitor} and \eqref{xitor} only depend  on $r^2$ and $\eta^2$,
therefore one should consider the latter as the actual natural variables for the bulk channel decomposition. Accordingly, we will often see $\partial_{r^2}$ and $\partial_{\eta^2}$ in the following formulae.

\section{The conformal blocks in radial coordinates}
\label{sec:scalar}

One of the advantages of the $\rho$ coordinates is that they have a clean geometric interpretation. In this section, we show how to exploit this fact in the computation of conformal blocks. For illustrative purposes, in this work we focus on the correlator of external scalar primaries \eqref{newStructEmi}. Nevertheless, we would like to emphasize that the same techniques can be used to tackle correlation functions of operators which carry non-trivial representations of $SO(d)$. This is the object of a forthcoming publication \cite{Lauria:2018klo}.

\subsection{Bulk Channel }
The bulk channel partial wave decomposition of the two-point function \eqref{newStructEmi} can be written as follows:
\begin{align}
\begin{split}
\label{scalarCPWbulk}
\langle \Ocal_1(P_1) \Ocal_2(P_2)\rangle_D &=\sum_{\Ocal}  c_{12\Ocal} a_{\Ocal} \ G_{\D,l}(P_1,P_2)
\\
&=
\sum_{\Ocal}  c_{12\Ocal} a_{\Ocal}
 \raisebox{1.8em}{
$
\xymatrix@=1.7pt{
{\Ocal_{1}}\ar@{-}[rdd]& &&&{ \ar@{=}[dd]  }\\  
&&&&\\
& *+[o][F]{}  \ar@{.}[rrr]^{\Ocal } & && *+[o][F]{} \ar@{=}[dd]  \\
&&&\\
\Ocal_{2} \ar@{-}[ruu]&&&& }
$
}
\end{split}
\end{align} 
Recall that $\Ocal_i$ are scalar primary operators with dimension $\D_i$, while $\Ocal$ is a primary operator with dimension and spin $\D,l$ respectively. Only traceless and symmetric tensors are exchanged by scalar external primaries. The OPE data $c_{12\Ocal}$ and $a_{\Ocal}$ appear in the three point function $\braket{\Ocal_1\Ocal_2\Ocal_{\Delta,l}}$ and in the one-point function $\braket{\Ocal_{\Delta,l}}_D$ respectively. We report them in appendix \ref{app:conventions}, with our choice of normalizations.
The quadratic Casimir gives rise to a second order differential equation:
\be
\label{Casimir_Scalar_Bulk}
\frac{1}{2}(J_1+J_2)^2 G_{\D l}(P_1,P_2) = -c_{\D l} G_{\D l}(P_1,P_2) \ ,
\ee
where the eigenvalue is $c_{\D l} =\D(\D-d) + l (l+d-2)$ and the generators of conformal transformations are $J_i^{M N} \equiv P_i^M \partial_{P_i}^N- P_i^N \partial_{P_i}^M $.
The explicit form of the equation is presented in appendix \ref{app:scalarBulk}. 
A general solution is not known in a closed form. In the following we present two techniques to obtain the partial waves $G_{\D l}$ as an expansion in the radial coordinates introduced in subsection \ref{subsec:rho}.
As a first step, we write the partial wave in terms of a function $g_{\D l}$ of the two cross ratios $r$ and $\eta$
\be
\label{GtoGcal}
G_{\D l}(P_1,P_2)\equiv\frac{\Acal(r,\eta)}{(P_1 \wbullet P_1 )^{\frac{\D_1}{2}} (P_2 \wbullet P_2 )^{\frac{\D_2}{2}}} \  g_{\D l}(r,\eta)\  .
\ee
For convenience, we stripped off the  function
\begin{align}
\label{S_prefactor}
   \Acal(r,\eta) \equiv (2 r)^{-\Delta _1-\Delta _2} \left(r^4-4 \eta ^2 r^2+2 r^2+1\right)^{\frac{1}{2} \left(\Delta _1+\Delta _2\right)} \ .
\end{align}
With this definition the function $g_{\D l}$ depends on the dimensions of the external operators only through their difference $\D_{12}=\D_1-\D_2$. 
Moreover, the partial wave $G_{\D,l}$ and the conformal block $ g_{\D,l}$ are simply related when considered in the bulk radial frame 
\be
\label{GrfGcal}
 G_{\D,l}(P_1,P_2)  \underset{b.r.f.}{\longrightarrow}  (2 r)^{-\Delta _1-\Delta _2}  \ g_{\D,l}(r, \eta ) \ ,
\ee
where the operation $ \underset{b.r.f.}{\longrightarrow}$ means that we set the points $P_1$ and $P_2$ to the bulk radial frame \eqref{B_Poin_Conf}.
Notice that the factor $(2 r)^{-\Delta _1-\Delta _2} $ takes into account the flat space scaling transformation $\l^{D} \Ocal(x) \l^{-D}=\l^{\D} \Ocal(\l x)$, $D$ being the generator of dilatations.
In fact, this factor is absent in the cylinder frame \eqref{PiOnTheCylinder1}.

We now show how to get $g_{\D,l}$ as an expansion in $r$. 
First we explain the meaning of this expansion in terms of the OPE. 
Then we obtain two recurrence relations which fix the coefficients of the expansion, following the ideas of \cite{Hogervorst:2013sma, radial_expansion}, \cite{Kos:2013tga,recrel}. 
The first one follows directly from the Casimir differential operator. The second one is derived by studying the pole structure in $\D$ of the conformal blocks.
We include in the submission a Mathematica file which computes the conformal blocks to a given order in $r$ following both strategies.

\subsubsection{A natural expansion }
\label{BulkNaturalExpansionScalar}
Conformal blocks admit a natural expansion in radial coordinates, where each power of $r$ measures the energy (on the cylinder) of some exchanged states in the conformal multiplet. In order to see this in detail, we define the conformal blocks in radial quantization:\footnote{To be precise, eq. \eqref{Gcalpq} is true up to a numerical factor, coming from the OPE data in eq. \eqref{scalarCPWbulk} and the factor 2 in eq. \eqref{GrfGcal}.} 
\be \label{Gcalpq}
g_{\D l}(r,\eta)
=
 \langle \hat 0 | 
 r^{H_{cyl}} \mathcal{P}_{\D,l} \
\mathcal{O}_1(n) \mathcal{O}_2(-n)
| 0
\rangle
 \ ,
\ee
where  $\mathcal{P}_{\D,l}$ is the projector onto the conformal family with highest weight labelled by $\D$ and $l$.
We used the Hamiltonian on the cylinder $H_{cyl}$ (a.k.a. the dilatation operator on the plane) to evolve the operators from the cylinder time $\t$, to the time $0$.

In order to diagonalize the action of $H_{cyl}$, it  is natural to write the projector as a sum over a complete basis of bulk states. Equation \eqref{Gcalpq} then becomes
\begin{align} \label{GcalSumStates}
&g_{\D,l}(r,\eta) =  \sum_{m=0}^\infty r^{\Delta+m}
\sum_{j=l-m}^{l+m}
\sum_{\dd} 
   \langle  \hat 0 | m, { \scriptsize
\ytableausetup{centertableaux,boxsize=1.2 em}
\begin{ytableau}
\m_1&\m_2&\, _{\cdots}&\m_j \\
\end{ytableau}
},\dd 
\rangle
\langle
m, { \scriptsize
\ytableausetup{centertableaux,boxsize=1.2 em}
\begin{ytableau}
\m_1&\m_2&\, _{\cdots}&\m_j \\
\end{ytableau}
},
\dd |
\mathcal{O}_1(n) \mathcal{O}_2(-n)
 \rangle
\,,
\end{align}
where we sum over all states at level $m$ of the conformal family, organized in irreducible representations (irreps) with spin $j$ of $SO(d)$. The index $\dd$ labels the degeneracy of such states.
The right overlap is  fixed by Lorentz symmetry 
\ytableausetup{centertableaux,boxsize=1.2 em}
\be
\langle
m, { \scriptsize
\ytableausetup{centertableaux,boxsize=1.2 em}
\begin{ytableau}
\m_1&\m_2&\, _{\cdots}&\m_j \\
\end{ytableau}
},
\dd |
\mathcal{O}_1(n) \mathcal{O}_2(-n)
 \rangle = 	
 u(m,j,\dd) \  n^{(\m_1} \cdots n^{\m_j)}
\ ,
\label{udefoverlap}
\ee
up to the coefficient $u(m,j,\dd)$. In \eqref{udefoverlap} and in the following, the parenthesis stands for symmetrization  and subtraction of the traces.
The  left overlap is fixed in terms of the following structure
\ytableausetup{centertableaux,boxsize=1.2 em}
\be
\label{One_Point_Radial_Frame}
  \langle  \hat 0 | m, { \scriptsize
\ytableausetup{centertableaux,boxsize=1.2 em}
\begin{ytableau}
\m_1&\m_2&\, _{\cdots}&\m_j \\
\end{ytableau}
},\dd 
\rangle  = 	
v(m,j,\dd) \ \pp^{(\m_1 \, \m_2} \cdots \pp^{\m_{j-1} \, \m_j)} \ .
\ee
Here we are allowed to use the projector $ \pp$ -- which is here the projector for spherical defects, according to eq. \eqref{ProjectorsRealSpace} -- because the overlap is computed in the vacuum of the defect theory. Notice that the angular cross  ratio is defined by $\eta^2 \equiv n \cdot \pp \cdot n$.
In formula \eqref{One_Point_Radial_Frame} it is clear that the indices $\m_1 \dots \m_j$ need to appear in even number. From this simple argument we obtain that the exchanged operators in a two-point function of scalars are in traceless and symmetric representations with even spins.\footnote{On the other hand, when we fix $d$ and $q$, we can use the epsilon tensor to write non trivial one-point functions for operators in more complicated $SO(d)$ representations \cite{Billo:2016cpy}. In particular, when $q=2$, symmetric pseudo-tensors with odd spin may be exchanged in the OPE. Here we only consider representations that are exchanged generically.}
\\
Combining \eqref{GcalSumStates} with \eqref{udefoverlap} and \eqref{One_Point_Radial_Frame} we obtain the following expansion
\be
\label{Bulk:RadialExpansion}
g_{\D l}(r,\eta)=r^{\D}  \ \sum_{m\geq 0} \ \sum_{j=\max[l-m,0]}^{l+m} \ w({m,j}) \  \mathfrak{F}_{m,j}(r,\eta) \ .
\ee
Here, $w({m,j})=\sum_\dd u(m,j,\dd)  v(m,j,\dd) $, while the basis of function $\mathfrak{F}_{m,j}(r,\eta)$ is defined as
\be
\mathfrak{F}_{m,j}(r,\eta)  \equiv r^m \mathfrak{C}_{j}(\eta) \ .
\ee
The function $\mathfrak{C}_{j}(\eta) $ plays in defect CFTs the role of the Gegenbauer polynomial in the ordinary four-point function. It is defined as
\ba
\label{DEF:Cj}
\begin{split}
\mathfrak{C}_{j}(\eta)  
& \equiv \pp^{(\m_1 \, \m_2} \cdots \pp^{\m_{j-1} \, \m_j)} \  n_{(\m_1} \cdots n_{\m_j)} \\
&= 
\frac{\left(\frac{1-j-p}{2} \right)_{\frac{j}{2}} }{\left(\frac{d+j-2}{2}\right)_{\frac{j}{2}}} \ \ \, _2F_1\left(-\frac{j}{2},\frac{d+j-2}{2};\frac{p+1}{2};\eta ^2\right)
 \ .
 \end{split}
\ea
By applying the Casimir of $SO(d)$ to eq. \eqref{udefoverlap}, one readily finds that
\beq
\nabla_\mu\nabla^\mu \mathfrak{C}_{j}(\eta) = -j(j+d-2)\mathfrak{C}_{j}(\eta)~,
\qquad
\nabla_\mu = \frac{\partial}{\partial n^\m} - n_\m\, n \cdot \frac{\partial}{\partial n}~,
\eeq
$\nabla_\mu\nabla^\mu$ being the Laplacian on the unit sphere.
 For concreteness we table the functions for the first few values of $j$.
\be
\begin{array}{|c|l|}
\hline 
j& \;\; \mathfrak{C}_{j}(\eta)    \phantom{\Big( }\\
\hline 
 \;\; 0 \;\; &\;\; 1  \phantom{\Big( } \\
 2 &\;\; \eta ^2-\frac{p+1}{d}   \phantom{\Big( }\\
 4 &\;\; \eta ^4-\frac{2 (p+3) \eta ^2}{d+4}+\frac{(p+1) (p+3)}{(d+2) (d+4)}  \phantom{\Big( } \\
 6 &\;\; \eta ^6-\frac{3 (p+5) \eta ^4}{d+8}+\frac{3 (p+3) (p+5) \eta ^2}{(d+6) (d+8)}-\frac{(p+1) (p+3) (p+5)}{(d+4) (d+6) (d+8)}  \phantom{\Big( } \\
\hline
\end{array}
\ee
As it is clear from the table, the prefactors in \eqref{DEF:Cj} normalize the coefficient of the highest power of the polynomial to one.

The convenience of the expansion \eqref{Bulk:RadialExpansion} stems from the fact that at each level $m$ finitely many coefficients appear. Specifically, at level $m$ there are at most $2 m+1$ unknowns.\footnote{For odd $m$ all the coefficients $w({m,j})$ are zero: the expansion could be written in terms of a powers series of $r^2$. See the comment after eq. \eqref{One_Point_Radial_Frame}.}  The level $m=0$ sets the overall normalization of the conformal block, which can be fixed arbitrarily. 
For later convenience, we choose $w({0,l})=4^\D$.
This corresponds to setting the leading OPE limit (small $r$ limit) of the conformal block to
\be
\label{LeadingOPEBulkScalarCB}
g_{\D l}(r,\eta)=(4r)^{\D} \mathfrak{C}_{l}(\eta) \  [1 + O(r^2)] \ .
\ee

As advertised, we now explain two strategies to fix the coefficients in the radial expansion \eqref{Bulk:RadialExpansion}.

\subsubsection{Casimir recurrence relation} 
\label{BulkBruteForceScalar}
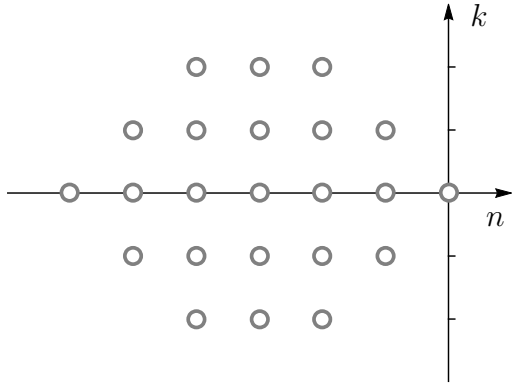
\begin{figure}[t!]
\graphicspath{{Fig/}}
\def\svgwidth{7 cm}
\centering
\input{scalarrec.pdf_tex} 
\caption{\label{ScalarRec} The set of points $\mathcal S$ in the recurrence relation for the coefficients $w(m+n,j+k)$.}
\end{figure}

From the Casimir differential equation \eqref{Casimir_Scalar_Bulk}, or better from its explicit form in radial coordinates \eqref{CasimirScalarBulk}, one easily obtains a recurrence relation for the coefficients $w(m,j)$.
First, we classify the action of a set of simple building blocks - we choose $r, \partial_r, \eta^2$ and $\partial_{\eta^2}$ - on the basis $ \mathfrak{F}_{m,j}$,
\be
\label{Scalar_buildingAction}
\left\{
\begin{array}{cl}
r              &\mathfrak{F}_{m,j}(r,\eta)= \mathfrak{F}_{m+1,j}(r,\eta) \\
\partial_{r} & \mathfrak{F}_{m,j}(r,\eta) =  \mathfrak{F}_{m-1,j}(r,\eta) \\
\eta^2      & \mathfrak{F}_{m,j}(r,\eta)=
 \mathfrak{a}_j \mathfrak{F}_{m,j-2}(r,\eta)+ 
 \mathfrak{b}_j \mathfrak{F}_{m,j}(r,\eta)+
  \mathfrak{F}_{m,j+2}(r,\eta) \\
\partial_{\eta^2} & \mathfrak{F}_{m,j}(r,\eta) =\frac{1}{(\eta^2 -1) \eta^2} \left[
 \mathfrak{c}_j \mathfrak{F}_{m,j-2}(r,\eta)+ 
\mathfrak{d}_j \mathfrak{F}_{m,j}(r,\eta)+
\frac{j}{2} \mathfrak{F}_{m,j+2}(r,\eta)
\right]
\end{array}
\right. \ .
\ee
Here we defined 
\be
\label{Scalar_buildingAction_2}
\begin{array}{llll}
\mathfrak{a}_j=
&
\frac{j (d+j-4) (j+q-3) (d+j-q-1)}{(d+2 j-6) (d+2 j-4)^2 (d+2 j-2)}
&
 \qquad 
 \mathfrak{c}_j=
&\frac{2 - d - l}{2} \ \mathfrak{a}_j 
\\
 \mathfrak{b}_{j}=
 & 
\frac{d^2+d (2 j-q-3)+2 \left(j^2-2 j+2 q-2\right)}{(d+2 j-4) (d+2 j)}
&
 \qquad 
\mathfrak{d}_j=
&
\frac{j (d+j-2) (d-2 q+2)}{2 (d+2 j-4) (d+2 j)}
\\
\end{array}
\ .
\ee
Eqs. (\ref{Scalar_buildingAction}-\ref{Scalar_buildingAction_2}) allow to 
recast any polynomial differential equation - by which we mean that the building blocks $r, \partial_r, \eta^2$ and $\partial_{\eta^2}$ appear polynomially - as a linear equation for the functions $\mathfrak{F}_{m,j}$ with shifted labels $m$ and $j$. Actually, $\partial_{\eta^2}$ is special since it also divides the right hand side by $(\eta^2 -1) \eta^2$. 
However one can always multiply the full differential equation by powers of $(\eta^2 -1) \eta^2$, thus obtaining the sought linear equation. 
Applied to the Casimir equation \eqref{CasimirScalarBulk}, the procedure yields a linear relation for the coefficients $w(m,j)$, which has the following schematic form 
\be
\sum_{(n, k) \in \mathcal S} c_{n,k} \ w(m+n, j+ k)= 0 \ ,
\label{eq:Scalar_recurrence}
\ee
where $c_{n,k} $ are coefficients that can depend on $\D,l,m,j,d,p$ and $\D_{12}$. The set $\mathcal S$ is a finite dimensional set of points, which represents all the possible shifts. We depict $\mathcal S$ in figure \ref{ScalarRec}. 
Notice that one can use eq. \eqref{eq:Scalar_recurrence} to express the right most point in terms of a linear combination of all the points on the left. Therefore eq. \eqref{eq:Scalar_recurrence} is a recurrence relation for the coefficient  $w(m,j)$, from which one can iteratively construct the function $g_{\D l}(r,\eta)$ at an arbitrarily high order in the radial expansion \eqref{Bulk:RadialExpansion}.
One can easily check that the obtained expansion (at any given order) solves exactly the Casimir equation (at that order).

We included in the submission a Mathematica notebook that contains the explicit equation \eqref{eq:Scalar_recurrence}. Therein we also define a function which builds the conformal blocks at any given order in $r$ according to equation \eqref{Bulk:RadialExpansion}. Finally we explicitly check that the computed conformal blocks satisfy the Casimir equation.

\subsubsection{Zamolodchikov recurrence relation } 
\label{BulkZamolodchikovScalar}
In this subsection we obtain a recurrence relation directly for the conformal blocks $g_{\D l}$ defined in \ref{BulkNaturalExpansionScalar}. The main idea was introduced by Zamolodchikov \cite{Zamolodchikov:1985ie} to study $d=2$ Virasoro blocks. Generalizations to higher dimensions can be found in \cite{Kos:2013tga,Kos:2014bka,recrel,radial_expansion,Isachenkov:2017qgn}. 
The conformal block can be written as a sum over all the poles in $\D$ and the analytic part. Since the residue at each pole has to be a new conformal block with different labels $\D$ and $l$ (this is proven in odd dimensions in \cite{recrel}), this expansion provides a natural recursive formula to compute conformal blocks. We now explain how to apply the same strategy to the bulk channel conformal block in a defect CFT.

The first step is to express the partial wave as a sum over states in radial quantization, in the same spirit of eq. \eqref{GcalSumStates}:
\be \label{CBRadialQuantization}
a_{\Ocal}c_{12\Ocal} G_{\D l}(P_1,P_2) = \sum_{\a \in \Hcal_{\Ocal}} \frac{ \langle \hat0 | \a \rangle \langle \a | \Ocal_1(P_1)\Ocal_2(P_2) | 0 \rangle}{\langle \a | \a \rangle} \ ,
\ee
where $\Hcal_{\Ocal}$ is the conformal multiplet associated to the primary $\Ocal$. 
Using the notation of \cite{recrel}, we find that at special values $\D=\D^\star_A$ in the multiplet $ \Hcal_{\Ocal}$ there is a descendant state $|\Ocal_A \rangle$ (with dimension $\D_A = \D^\star_A+n_A $ and spin $l_A$), which becomes primary. When this special event occurs, the multiplet $ \Hcal_{\Ocal}$ becomes reducible, and a new irreducible submultiplet $ \Hcal_{\Ocal_A}$ of null states breaks off. Because of the null states, the conformal block \eqref{CBRadialQuantization} has to diverge at $\D=\D^\star_A$. In fact, it is possible to prove \cite{recrel} that in odd dimension the divergence is just a single pole\footnote{For odd spacetime dimensions it is proved that the conformal blocks only have single poles in $\D$. For even dimensions there can exist higher order poles. However the Zamolodchikov recurrence relation provides a good analytic continuation in dimensions, which still holds in the limit of even dimension.} which has residue proportional to the CB labeled by the primary descendant $\Ocal_A$,
\be
G_{\D l}(P_1,P_2) =  \frac{R_A}{\D-\D^\star_A}  G_{ \D_A  l_A}(P_1,P_2) +O((\D-\D^\star_A)^0) \ .
\label{eq:GDeltal1}
\ee
The coefficient $R_A$ can be computed explicitly. It accounts for the different normalization of the primary descendant with respect to the one used for primary operators.
In particular, the two point functions $\langle \Ocal_A \Ocal_A \rangle$, the three point functions $\langle \Ocal_1\Ocal_2\Ocal_A \rangle$ and the one point functions  $\langle \Ocal_A \rangle_D$ are not canonically normalized. Therefore we parametrize $R_A$ as follows:
 \be
 R_A=M_A^{(L)} Q_A M_A^{(R)} \ ,
 \label{QMM}
 \ee
where schematically we define
\begin{equation}\label{MbulkSca}
\begin{array}{cc c l}
 \langle \Ocal_A \rangle_D&= M_A^{(L)} 
&\langle \Ocal \rangle_D &\\
\langle \Ocal_1\Ocal_2\Ocal_A \rangle&= M_A^{(R)} & \langle \Ocal_1\Ocal_2 \Ocal \rangle&
\\
\langle \Ocal_A \Ocal_A \rangle^{-1}&= \frac{Q_A}{\D-\D^\star_A} &\langle \Ocal \Ocal \rangle^{-1} &+ O((\D-\D^\star_A) ^0)
\end{array}
\ ,
\end{equation}
and $\Ocal$ is a canonically normalized primary operator with the same quantum numbers as $\Ocal_A$.
The pole structure of the conformal blocks in the bulk channel is identical to the one of a conformal block for a four-point function of local operators. This is not surprising, since the norm $\langle \Ocal_A \Ocal_A \rangle$, which governs the position of the poles, does not know of the presence of the defect. For the same reason, the coefficients $M_A^{(R)}$ and $Q_A$ are the same as the ones computed in \cite{recrel}.\footnote{Our choice of conventions implies that $M_A^{(R)}$ corresponds to $M_A^{(L)}$ in \cite{recrel}.} The only new ingredient is $M_A^{(L)}$, which is computed in appendix \ref{App:ZamolodchikovBulkScalar}.

We can now reconstruct the full conformal block by summing over all the poles in $\D$ and the regular part. To do so it is more convenient to use the  block $h_{\D l}=(4r)^{-\D}g_{\D l}$ which has a regular limit for large $\D$. The regular part, which we call $h_{\infty l}$, is computed in eq. \eqref{BulkLargeDeltaScalar} in appendix \ref{app:scalarBulk} by solving the Casimir equation at large $\D$.
The final result is as follows, 
\begin{align}
h_{\D l}(r,\eta)&=h_{\infty l}(r,\eta)+\sum_{A} \frac{R_A }{\D-\D^\star_A} (4 r)^{n_A}
h_{\D_A\,l_A}(r,\eta) \ .
\label{eq:recscalar}
\end{align}
All the ingredients of this formula have been defined previously in this subsection. The only missing information is the set of labels $\D^\star_A,\D_A,l_A,n_A$, which again matches what was already found in \cite{recrel}. The label $A$ stands for two indices: a type index $T=\I,\II,\III$ and a natural number $n$. For each type, the natural number $n$ takes value into different sets which may be infinite (for type $\I$ and $\III$) or finite (for type $\II$). We give the complete set of labels  in the following table:
\be\label{tableDelta}
\begin{array}{|c | ccc|}
\hline
\phantom{\Big(}		A														&\D^\star_A 					&n_A 										&l_A	 \\ 
\hline
\phantom{\Big(} \mbox{Type \I: }  n=2,4,6,\dots 	\infty					& 1-l-n 			 				& n 	  	   									 &l + n \\ 
\phantom{\Big(} \quad  \mbox{Type \II: }n=2,4,6,\dots,l \;\; \quad 	&\quad   l+2h-1-n  	\quad & \quad n \quad  						&\quad l - n \quad  \\
\phantom{\Big(} \mbox{Type \III: }n=1,2,3\dots \infty \; 				& h-n  							& 2n       			 						 &l       \\ 
\hline
\end{array}
\ee
Notice that the values of $n_A$ in the table are only even integers. Odd values would give rise to odd powers in $r$ in the conformal blocks, which were excluded in subsection \ref{BulkNaturalExpansionScalar}.

We would like to remark that \eqref{eq:recscalar} can be efficiently used to compute the conformal blocks in a radial expansion. In fact, this equation should be interpreted as a recurrence relation for $h_{\D l}$ where the term $h_{\infty l}$ is a seed. Each time we iterate \eqref{eq:recscalar} we are ensured to obtain higher order contributions in the power series in $r$, because the numbers $n_A$ are positive. This kind of recurrence relations is in fact the standard technique to compute conformal blocks for the numerical bootstrap of a four-point function of local operators.

More details about the recurrence relation are described in appendix \ref{App:ZamolodchikovBulkScalar}, where we review how to construct the primary descendant states $\Ocal_A$ and we compute the coefficients $R_A$.

All the definitions  that enter in formula \eqref{eq:recscalar} are summarized in a Mathematica notebook which we include in the submission.
A function which computes the conformal blocks $h_{\D l}$ at any given order in $r$ using \eqref{eq:recscalar} is also defined to the notebook.
Finally two checks are added: first, that the computed blocks satisfy the Casimir equation and second, that they are equal to the blocks generated with the strategy proposed in sections \ref{BulkNaturalExpansionScalar} and  \ref{BulkBruteForceScalar}.

\subsection{Defect Channel}
We now focus on the defect conformal block decomposition of the two-point function \eqref{newStructEmi}:
\ba
\label{scalarCPWdef}
\begin{split}
\left\langle 
\mathcal{O}_1\left(P_1\right) \mathcal{O}_2\left(P_2\right)
\right\rangle_D
&=\sum_{\hO} b_{1 \hO} \, b_{2 \hO} \, \hG_{\hO}(P_1,P_2) \\
&=
\sum_{\hO}  b_{1\hO} b_{2 \hO}
 \raisebox{2.6em}{
$
\xymatrix@=1.6pt{
&&& \ar@{=}[dd]  \\
&&& \\
{\Ocal_{1}}\ar@{-}[rrr]& &&*+[o][F]{} \ar@{.}[dd]  \\  
&  &&
\\
\Ocal_{2} \ar@{-}[rrr]&&&*+[o][F]{} \ar@{=}[dd] \\
&&& \\
&&& }
$
}
\mbox{\scriptsize$\hO$} \ .
\end{split}
\ea
The defect operators $\hO$ are labeled by their conformal dimension $\hD$ and their parallel and perpendicular spins $\hl,s$. Since the external  operators $\Ocal_i$ are scalars the operators $\hO$ are restricted to have $\hl=0$. The OPE data $b_{i\hO}$ appear in the two-point function $\braket{\Ocal_i\hO}_D$. We report its form in appendix \ref{app:conventions}, with our choice of normalizations.

The defect conformal partial waves $ \hG_{\hO}(P_1,P_2)$ are eigenfunctions of the quadratic Casimir of the defect group $SO(p+1,1)\times SO(q)$, which factorizes in a parallel and a transverse part. We therefore obtain two differential equations:
\ba
 \frac{1}{2}\left(J_1^\pdot\right)^2  \hG_{\hO}(P_1,P_2) &=& -c^\pdot_{\hO}  \hG_{\hO}(P_1,P_2) \ , \\
\frac{1}{2}\left(J_1^\tdot\right)^2  \hG_{\hO}(P_1,P_2) &=& -c^\tdot_{\hO}  \hG_{\hO}(P_1,P_2) \ ,
\ea
where $J_1^{M N} \equiv P_1^M \partial_{P_1}^N- P_1^N \partial_{P_1}^M $ and the suffix $\pdot$ ($\tdot$) means that the indices are first projected onto the parallel (orthogonal) space via $\Pt$ ($\Pp$) and then contracted. The eigenvalues associated to an operator $\hO$ labelled by $\hD,\hl=0,s$ are
\be
c^{\pdot}_{\hD \hl=0} =\hD(\hD-p) \ ,
\qquad
c^{\tdot}_{s} =s (s+q-2) \ .
\ee

It is convenient to write the partial waves  $\hG_{\hO}$ in terms of conformal blocks $\hg_{\hO}$  which only depend only on the cross ratios $\hr$, $\heta\,$:
\be
\label{GtoGcal_D}
\hG_{\hO}(P_1,P_2) \equiv \frac{1}{(P_1 \wbullet P_1 )^{\frac{\D_1}{2}} (P_2 \wbullet P_2 )^{\frac{\D_2}{2}}}\ \hg_{\hO}(\hr,\heta)\  .
\ee
Notice that $\hG_{\hO}$ is simply related to $\hg_{\hO}$ when computed in the defect radial frame \eqref{D_Poin_Conf}
\be
\hG_{\hO}(P_1,P_2)  \underset{d.r.f.}{\longrightarrow} \frac{1}{\hr^{\D_2}}\,\hg_{\hO}(\hr,\heta)  \ .
\ee
Here $ \underset{d.r.f.}{\longrightarrow}$ means that the points $P_i$ are set in the radial frame \eqref{D_Poin_Conf}.

The scalar conformal block in the defect channel was computed in a closed form in \cite{Billo:2016cpy}  by solving the Casimir equation.
As consequence of the factorization of the Casimir, it also nicely factorizes:
\be
\label{D_CB_scalar}
\hg_{\hD,s}(\hr,\heta)=\Wcal
(\hr)\ \Ccal_{s}(\heta) \ ,
\ee
where 
\be
\label{CB_Rad_Ang}
\Wcal
(\hr) \equiv \hr^{\hD } \ _2F_1\left(\frac{p}{2},\hD ;\hD -\frac{p}{2}+1; \hr ^2\right) \ , \qquad 
\Ccal_{s}(\heta) \equiv 
\frac{s!}{2^{s}\left(\frac{q}{2}-1\right)_s}  
C_s^{\frac{q}{2}-1}(\heta)
\ ,
\ee
and $C_n^{\n}(x)$ is a Gegenbauer polynomial. It is worth to notice that in radial coordinates the conformal block is particularly simple. 
In fact for even $p$, the hypergeometric  function in $\Wcal$ reduces to a rational function of $\hr$. For example for $p=2$, $\, _2F_1 (1,\hD ;\hD; \hr^2)=\frac{1}{1-\hr^2}$.

Even if the blocks are known exactly, in the following we show how to obtain them in two alternative ways. This allows to clarify the solution \eqref{CB_Rad_Ang} in light of the defect OPE, included the analytic structure in $\hD$. Furthermore, it is a useful warm for the case of external operators with spin, which we shall address in a forthcoming paper \cite{Lauria:2018klo}.

\subsubsection{A Natural Expansion}
\label{natural_expansion_defect_scalar}
In this subsection we show that the expansion in $\hr$ descends naturally from the OPE, where the power of $\hr$ measures the conformal dimension of the exchanged states in the conformal multiplet. Since the transverse part of the defect group only consists of rotations, Lorentz invariance alone determines the function of $\heta$. The coefficients of the $\hr$ expansion can instead be fixed using a recurrence relation that descends from the Casimir equation.

We begin by writing the conformal blocks defined in eq. \eqref{GtoGcal_D} in radial quantization: 
\be 
\hg_{\hD s}(\hr,\heta)
=
 \left\langle \hat 0 |
\mathcal{O}_1\left(n\right)
\hr^{H_{cyl}} \mathcal{P}_{\hD s}
\mathcal{O}_2\left(n'\right)
|\hat 0  \right\rangle
 \ ,
 \label{Gcalpq_D}
\ee
where $\hD$ and $s$ are the dimension and transverse spin of the exchanged defect primary $\hOcal$, and $\mathcal{P}_{\hD s}$ projects onto its multiplet. In terms of a complete set of states,
\begin{align} 
&\hg_{\hD s} =  \sum_{m=0}^\infty r^{\hD+m}
 \langle \hat 0 |
\mathcal{O}_1(n)
| m, { \scriptsize
\ytableausetup{centertableaux,boxsize=1.2 em}
\begin{ytableau}
i_1&\, _{\cdots}&i_s \\
\end{ytableau}
}
  \rangle
   \langle m, { \scriptsize
\begin{ytableau}
i_1&\, _{\cdots}&i_s \\
\end{ytableau}
}
|
\mathcal{O}_2(n')
| \hat 0
\rangle
\,.
\label{GcalSumStates_D}
\end{align}
The indices $i_k$ belong to the transverse space and accordingly the states $| m, { \scriptsize
\ytableausetup{centertableaux,boxsize=1.2 em}
\begin{ytableau}
i_1&i_2&\, _{\cdots}&i_s \\
\end{ytableau}
}
 \rangle
$
are in a traceless and symmetric representation of $SO(q)$.
Notice that in this case each descendant at level $m$ is unique, therefore we can omit the label $\dd$ of the degeneracy.
The overlap are fixed by Lorentz symmetry to take the following form
\ytableausetup{centertableaux,boxsize=1.2 em}
\be
\label{2Bd:0j}
\langle 
\hat 0 | 
\mathcal{O}_1(n)
| m, { \scriptsize
\ytableausetup{centertableaux,boxsize=1.2 em}
\begin{ytableau}
i_1&i_2&\, _{\cdots}&i_s \\
\end{ytableau}
}
 \rangle
 = 
 u(m) \ n^{(i_1}\dots n^{i_s)} \, ,
\ee
where $u(m)$ are numerical coefficients.
Here the parenthesis in $n^{(i_1}\dots n^{i_s)}$ implement symmetry and tracelessness for $SO(q)$ representations.
Multiplying the left and right overlaps we automatically obtain the angular part of the conformal block
\be
 n^{(i_1}\dots n^{i_s)}  n'^{(i_1}\dots n'^{i_s)} = \Ccal_{s}(\heta)  \ ,
\ee
as defined in \eqref{CB_Rad_Ang}.

We then obtain the natural expansion 
\begin{align} 
\label{D_Ansatz_BF_scalar}
&\hg_{\hD s}(\hr,\heta) =  \Ccal_{s}(\heta) \ \Wcal(\hr) \, , \qquad  
\Wcal(\hr)\equiv \sum_{m=0}^\infty 
w(m) \ \hr^{\hD+m} \ ,
\end{align}
where
$
w(m)=  u(m) \tilde u(m)
$ and $\tilde u(m)$ is the coefficient of the right overlap. This ansatz solves the perpendicular part of the Casimir equation.
The parallel part acts on $\Wcal(\hr)$ giving rise to a recurrence relation for the coefficients $w(m)$,
\be
\label{Rec_Rel_w_Scalar_Defect}
(2 \Delta +m-1) (m+p-1) \ w(m-1)-(m+1)  (2 \Delta +m-p+1) \ w(m+1)=0 \ .
\ee
This recurrence relation has a unique solution once we fix the initial condition $w(-1)=0$, $w(0)=1$ (the value of $w(0)$ sets the normalization of the conformal blocks),
\be
\label{D_w_scalar}
w(2m)=\frac{\left(\frac{p}{2}\right)_m (\hat \Delta )_m}{m! \left(-\frac{p}{2}+\hat \Delta +1\right)_m} \ , \qquad
w(2m+1)=0 \ .
\ee
The series in equation \eqref{D_Ansatz_BF_scalar} can then be resummed, to find the conformal blocks in a closed form as in  \eqref{D_CB_scalar}.

\subsubsection{Zamolodchikov recurrence relation} 
\label{Zamol_Defect_Scalar}
We now explain how to develop a recurrence relation for the defect conformal blocks by studying their analytic properties in the variable $\hD$.
We first write the conformal block  as a sum over states in radial quantization 
\be \label{CBRadialQuantization_D}
b_{1\Ocal}b_{2\Ocal} \hG_{\hD s}(P_1,P_2) = \sum_{\a \in \hHcal_{\hOcal}} \frac{ \langle \hat 0| \Ocal_1(P_1) | \a \rangle \langle \a | \Ocal_2(P_2)| \hat0 \rangle}{\langle \a | \a \rangle} \ ,
\ee
where $\hHcal_{\hOcal}$ is the conformal multiplet associated with the defect primary $\hOcal$. Similarly to the bulk case (see subsection \ref{BulkZamolodchikovScalar}), when $\hD=\hD^\star_A$ a descendant state $|\hOcal_A \rangle$ (with dimension $\hD_A = \hD^\star_A+n_A $ and transverse spin $s_A$) becomes primary, the representation $ \hHcal_{\hOcal}$ becomes reducible and we expect the following polar structure for the conformal blocks as functions of $\hD$,
\be
\hG_{\hD s}(P_1,P_2) =  \frac{\hR_A}{\hD-\hD^\star_A}  \hG_{ \hD_A  s_A}(P_1,P_2) +O((\hD-\hD^\star_A)^0) \ .
\label{eq:GDeltal1_D}
\ee
The coefficient $\hR_A$ can again be computed as
 \be
 \hR_A=\hM_A^{(L)} \hQ_A \hM_A^{(R)} \ ,
 \label{QMM_D}
 \ee
where 
$
\langle \Ocal_1 \hOcal_A \rangle_D= \hM_A^{(L)}  \langle \Ocal_1 \hOcal \rangle_D
$
and similarly for $ \hM_A^{(R)}$ (where $\hOcal$ is a canonically normalized primary defect operator with the same quantum numbers as $\hOcal_A$), while $\hQ_A$ comes from the inverse of the norm of the intermediate state $\alpha$.
Let us start by presenting the solution. The conformal block is reconstructed from a single tower of null descendants (type \III):\footnote{We name it type \III in order to match the bulk channel conventions.}
\begin{align}
\begin{split}
\label{D_CB_RecRel}
\hg_{\hD s}(\hr,\heta)&= (\hr)^{\hD} \hat h_{\hD}(\hr) \Ccal_{s}(\heta) \ ,
\\
\hat h_{\hD}(\hr)&=\hat h_{\infty}(\hr)+\sum_{n=1}^\infty \frac{\hR_{\III,n} (\hr)^{2n} }{\hD-(\frac{p}{2}-n)} 
\hat h_{\frac{p}{2}+n }(\hr) \ .
\end{split}
\end{align}
Here the angular part $\Ccal_{s}(\heta)$ defined in  \eqref{CB_Rad_Ang} is fixed by the leading OPE and is factorized. 
The regular part $\hat h_{\infty }$ is easily obtained by solving the parallel part of the Casimir equation at  leading order in large $\hD$, 
\be
\hat h_{\infty }(\hr) =\left(1-\hr^2 \right)^{-\frac{p}{2}} \ .
\ee
The coefficient of proportionality \eqref{QMM} is
\be
\label{RIII_D}
\hat R_{\III,n}\equiv \frac{(-1)^{n-1} \left(\frac{p}{2}-n\right)_{2 n}}{(n-1)! n!} \ .
\ee

Let us now explain how to get eqs. \eqref{D_CB_RecRel} and \eqref{RIII_D}. 
The simplicity of the recurrence relation stems from a constraint which is obvious from eq. \eqref{2Bd:0j}: only defect operators with vanishing parallel spin couple to a bulk scalar.
The form of the defect conformal multiplet is the one of a $p$-dimensional CFT: the null states are the same as in table \ref{tableDelta}. Everything regarding their definition and the computation of their norms was already addressed in  \cite{recrel} and can be simply used just replacing $d$ with $p$. But the mentioned constraint implies that the only coupled descendants are of the form $(P \pdot P)^n | \hat \Ocal\rangle$, for $n=1,\dots \infty$. 
 They become primaries when $\hat \D=p/2-n$, giving rise to the poles of the conformal blocks \eqref{D_CB_RecRel}. Since the primary descendants are at level $2n$, the blocks at the residue are labeled by $\hD=p/2+n$.

 The coefficient $\hR_{\III,n}\equiv \hQ_{\III,n} \hM_{\III,n}^2$ is obtained as follows.
$\hQ_{\III,n}$ is the same as in eq. \eqref{QA} once replaced $d\rightarrow p$ and set $l=0$. To compute $\hat{M}_{\III,n}$, we consider the two-point function
 \be
 \label{BD_scalar_flat}
 \langle \Ocal_1(y)  \hO^{i_1 \dots i_s}(x)\rangle_D= b_{\Ocal_1  \hO } \ y^{(i_1}\dots y^{i_s)} (y\tdot y)^{\frac{\hat \D-\D_1-s}{2}} (y\tdot y+x\pdot x)^{-2 \hat \D} \ ,
 \ee
 where the operator $\Ocal_1$ is at the origin of the parallel space, therefore $y$ only has transverse components (namely $y=\pt \cdot y$).
Taking derivatives of the operator $\hO$ we easily obtain $\hM_{\III,n}$, 
\begin{align}\label{scalarMIII}
\left(\, \partial_x \pdot \partial_x \, \right)^n (y\tdot y+x\pdot x)^{- (p/2-n)}&\equiv \hM_{\III,n} (y\tdot y+x\pdot x)^{- (p/2+n)} (y\tdot y)^{n} \ ,\\
\hM_{\III,n} &=(-4)^{n} \left(\frac{p}{2}-n\right)_{2n}  \ .
\end{align}

Notice that when $p$ is an even number, $\hR_{\III,n}$ evaluates identically to zero for any integer $n\geq p/2$. This truncation is the reason why the conformal blocks take a simple rational form when $p$ is even.

\section{OPE convergence in defect CFT}
\label{sec:convergence}

We would like to conclude this work with a discussion of the convergence of the defect and bulk OPEs. In particular, we begin by addressing two separate questions: the region of convergence of the sum over conformal blocks, and the radius of convergence of the expansion of the two-point function in powers of $r$ and $\hr$. Then, in subsection \ref{subsec:rate}, we refine the analysis and bound the rate of convergence of the defect OPE. We cannot do the same for the bulk OPE, due to the lack of positivity. Instead, in subsection \ref{subsec:landau} we compare the radial expansion of a single bulk channel block with the expansion in powers of $\xi$, the cross ratio defined in eq. \eqref{crossxizeta}. This further highlights the convenience of the cross ratios defined in this work.
The discussion will make use of standard technology, which can be found in \cite{Pappadopulo:2012jk}. 

Let us first discuss what is the region of convergence of the sum over conformal blocks. The standard way to prove convergence of an OPE relies on the completeness of a Hilbert space whose basis elements are in one to one correspondence with the scaling operators  of the theory \cite{Pappadopulo:2012jk}. This is naturally obtained in radial quantization. The OPE then converges whenever a sphere can be drawn which separates the insertions to be fused together from all the others. Different choices of the center of the sphere yield different convergent series expansions of the same conformal block. 
Indeed, these choices single out Hilbert spaces whose basis elements are related by translations. Hence, OPEs centered in different points differ by the size of the contribution of descendants. The conformal block is insensitive to this rearrangement. Similarly, defect local operators are in one-to-one correspondence with states on a sphere centered on the defect, and correspondingly the convergence of the defect OPE must be discussed using the latter.

In practice, since it is sufficient to establish convergence in the domain $\mathcal{D}$ or $\hat{\mathcal{D}}$  in eqs. \eqref{regD} and \eqref{regDhat}, the fastest approach is to discuss the configurations in figs. \ref{fig:rho}, \ref{fig:rhohat}. In order to fix ideas, we refer to the latter.
Let us first consider the bulk OPE. When $\hrho=\hrhobar<0$, the sphere that separates $\Ocal_1$ and $\Ocal_2$ degenerates. We conclude that the sum over bulk blocks converges everywhere in the fundamental region $\hat{\mathcal{D}}$ except on the negative real axis. 
Notice that the negative real axis is consistently mapped by eq. \eqref{rho(hrho)} to the boundary of $\mathcal{D}$ at $|\rho|=1$.
Let us now turn to the defect OPE. In this case, for every position $\hrho$ inside the unit circle there is a sphere centered on the defect in the origin which separates $\Ocal_2$ from $\Ocal_1$. Therefore, the defect OPE converges in the region $\hat{\mathcal{D}}$, except at $|\hrho|=1$. 

In the case of a codimension one defect, the transverse space is one-dimensional, but the previous considerations apply as well. If we name $y$ the transverse coordinate, and $\Ocal_1$ is placed at $y=1$, $\Ocal_2$ can be made to lie in the interval $0<y<1$ or $-1<y<0$. The bulk OPE converges in the former region, that is when $\Ocal_1$ and $\Ocal_2$ lie on the same side of the interface, while the defect OPE also converges when the interface separates them.

We now address the second question: the radius of convergence of the radial expansion of the two-point function. Since the techniques developed in sections \ref{sec:scalar} provide power series representations of the blocks, this question is clearly relevant. Again, the standard strategy is to rewrite the expansion as the decomposition of a vector of a Hilbert space into an orthogonal basis. This is easy to because the radial coordinates are conjugate to time evolution on a cylinder. Eqs. \eqref{GcalSumStates} and \eqref{GcalSumStates_D} define projections of the two following vectors over complete set of states: 
\beq
\ket{\Ocal_1\Ocal_2}=\Ocal_1(r n,z_1)\Ocal_2(-r n,z_2)\ket{0},\quad \textup{and} \quad\ket{\Ocal_2}=\Ocal_2(\hr n,z_1) \ket{\hat{0}}~,
\label{OOvec}
\eeq 
respectively. The projection is simply obtained summing eqs. \eqref{GcalSumStates} and \eqref{GcalSumStates_D} over the exchanged conformal families. The same equations also define the radial expansions of the two-point function: convergence of the expansions in $r$ ($\hr$) at fixed $\eta$ ($\heta$) is the same as convergence of the decomposition of $\ket{\Ocal_1\Ocal_2}$ and $\ket{\Ocal_2}$ in the bases  $\{\ket{\Delta, m, j,\dd}\}$  and $\{\ket{\hD,m,s}\}$ respectively, where with respect to eqs. \eqref{GcalSumStates} and \eqref{GcalSumStates_D} we added a label denoting the exchanged primary. Now, the convergence of the decompositions of finite norm states inside scalar products is a property of orthonormal bases in a Hilbert space, so convergence of the power series is guaranteed as long as $r<1$ ($\hr<1$). When $r=1$ ($\hr=1$), the norm of $\ket{\Ocal_1\Ocal_2}$ ($\ket{\Ocal_2}$) in the Hilbert space on the sphere of radius 1 diverges, and when $r>1$ ($\hr>1$) the two-point functions cannot be written in radial quantization in terms of the overlaps in eqs. \eqref{Gcalpq} and \eqref{Gcalpq_D}. 
We learn that the radii of convergence of the series expansion in $r$ and $\hr$ match the regions of convergence of the bulk and defect OPEs respectively. Clearly, we can repeat the previous considerations after projecting the vectors in eq. \eqref{OOvec} onto a single conformal family, so the $r$ and $\hr$ expansions of bulk and defect blocks converge in the same region. Analyticity of the defect blocks for $\hr<1$ can be checked explicitly in eq. \eqref{CB_Rad_Ang}, while in appendix \ref{app:d4q2} we check the analogous statement for the bulk blocks of a codimension two defect, which are known exactly. Of course, this analysis goes through unchanged in the case of external operators with spin.

We conclude the subsection with a remark on the bulk radial expansion of a two-point function. This expansion is never positive, but as all other OPE decompositions it can be easily proven that it is absolutely convergent.\footnote{We thank J. Penedones for a discussion on this point.} This is a generic property of the decomposition of a vector in a orthonormal basis. Indeed, consider again the decomposition of the vector $\ket{\Ocal_1\Ocal_2}$ in the basis $\{\ket{\Delta, m, j,\dd}\}$.
If we change arbitrarily the phase of each coefficient in the decomposition, we obtain another vector of equal, \emph{i.e.} finite, norm. Therefore convergence inside correlation functions is again guaranteed.

\subsection{Rate of convergence and the density of defect states.}
\label{subsec:rate}

In the bootstrap, an important question is the rate of convergence of the OPE representation of a correlation function. The answer depends of course on the correlation function itself, and on the choice of kinematics. In \cite{Pappadopulo:2012jk} -- see also \cite{Rychkov:2015lca} -- it was shown that the asymptotic rate of convergence of a four-point function of pairwise identical scalars is exponential, away from the boundary of the region of convergence of the OPE. This means that the role of high dimensional operators in ensuring crossing symmetry is limited. On one hand, this effective decoupling has been beneficial to the bootstrap. It implies that crossing symmetry constrains the low lying CFT data, rather than just linking the contribution of light operators in one channel and heavy operators in the other. More concretely, it puts the linear functional method \cite{arXiv:0807.0004} on solid mathematical ground, included the use of approximations for the conformal blocks, and it gives confidence in the results obtained with the determinant method \cite{Gliozzi:2013ysa}. On the other hand, the same decoupling makes it hard to extract information about high-dimensional operators. This issue can be partially overcome by combining the output of the numerics with analytic results obtained from the light-cone bootstrap \cite{Simmons-Duffin:2016wlq}.

Here we apply the method of \cite{Pappadopulo:2012jk} to the study of the defect OPE decomposition of a two-point function, and we focus for simplicity on the case of two identical scalar primaries. 
Along the way, we establish a result on the density of defect states weighted by the OPE coefficients, analogous to the one obtained in \cite{Pappadopulo:2012jk} for the ordinary bulk OPE. As we remark at the end of the subsection, we are unable to repeat the analysis for the bulk OPE decomposition.

As a first step, we notice that when $\hat{\eta}=1$ the expansion in powers of $\hr$ of the two-point function has positive coefficients. This follows from reflection positivity on the cylinder, and is obvious from eq. \eqref{GcalSumStates_D}, since $\heta=1$ implies $n=n'$. 
This will be important in a moment. We can conveniently express the expansion as a Laplace transform:
\beq
\braket{\Ocal(1,n)\Ocal(\hr,n)}_D=\int_0^\infty \!d E\, \hat{f}(E) \,e^{-\beta E},
\label{defLaplace}
\eeq
where $\beta=-\log\hr$. $\hat{f}(E)$ is recognized as a weighted density of defect states via eq. \eqref{GcalSumStates_D}:
\beq
\hat{f}(E) = \sum_{\hat{E}} |\bra{\hat{0}} \Ocal(1,n)\ket{\hat{E}}|^2 
\,\delta(E-\hat{E}).
\label{weightedf}
\eeq
Notice that $\hat{f}(E)$ does not depend on $n$.\footnote{To be precise, for each descendant of transverse spin $s$ we need to introduce the projector $\mathcal{P}_{\hat{E}}~=~| \hat{E}, { \scriptsize
\ytableausetup{centertableaux,boxsize=1.2 em}
\begin{ytableau}
i_1&\, _{\cdots}&i_s \\
\end{ytableau}
}
  \rangle
   \langle \hat{E}, { \scriptsize
\begin{ytableau}
i_1&\, _{\cdots}&i_s \\
\end{ytableau}
}
|$, where the transverse indices $i_k$ are summed over. The projector commutes with the transverse rotation generators, therefore in particular:
$\bra{\hat{0}} \Ocal(1,n) \mathcal{P}_{\hat{E}} \Ocal(1,n) \ket{\hat{0}}$ does not depend on the unit vector $n$.  \label{foot:projector}}
The asymptotic behavior of $\hat{f}(E)$ for large $E$ is controlled by the $\beta\to 0$ limit of the correlator, which in turn is dominated by the identity in the bulk channel:
\beq
\braket{\Ocal\Ocal}_D \overset{\beta\to 0}{\sim} (1-\hr)^{-2\Delta} 
\sim \beta^{-2\Delta}~,
\label{betaIdentity}
\eeq
where $\Delta$ is the scaling dimension of $\Ocal$. One can then use the Hardy-Littlewood Tauberian theorem -- see \cite{Pappadopulo:2012jk} -- to turn eq. \eqref{betaIdentity} into an asymptotic constraint for the integrated density defined as:
\beq
\hat{F}(E)=\int_0^E\! dE' \hat{f}(E').
\eeq
In particular,
\beq
\hat{F}(E) \overset{E\to \infty}{\sim} \frac{E^{2\D}}{\Gamma(2\D+1)}~,
\label{intdefdens}
\eeq
The theorem is valid for positive densities $\hat{f}(E)$, and the remarks before eq. \eqref{defLaplace} imply that this is the case. Without further assumptions, the subleading corrections to eq. \eqref{intdefdens} are only logarithmically suppressed ($O(1/\log E)$). As in \cite{Pappadopulo:2012jk}, it is instructive to compare the estimate \eqref{intdefdens} with the unweighted density of states $\hat{f}_0$, which enters the partition function of the theory on $S^{d-1}$ at finite temperature:
\beq
Z_{S^{d-1}}(\beta)=\int_0^\infty dE\, \hat{f}_0(E)\, e^{-\beta E},\qquad 
\hat{f}_0(E)=\!\!\sum_{\hat{E}\,\in\, \textup{spectrum}}\!\!\!\! \delta(E-\hat{E})~.
\label{Zsphere}
\eeq
As $\beta\to 0$, $Z_{S^{d-1}}(\beta)$ can be estimated via the flat space free energy density, which on dimensional grounds has the following behavior:
\beq
\lim_{\textup{Vol}\to\infty}\frac{1}{\textup{Vol}}\log Z_{\textup{Vol}}(\beta) = \frac{k_b}{\beta^{d-1}},\qquad k_b>0~.
\eeq
Here $Z_\textup{Vol}$ is the partition function on a flat finite geometry of size Vol. Positivity of $k_b$ follows from thermodynamic stability. Notice that the size of the defect, which in \eqref{Zsphere} marks the $S^{d-1}$ along an $S^{p-1}$ with the same radius, scales like the volume of the flat geometry. On dimensional ground, its contribution to $\log Z_{\textup{Vol}}$ scales like $\textup{Vol}^{(p-1)/(d-1)}$ and is therefore subleading. It follows that the high temperature limit of $Z_{S^{d-1}}$ is exponentially enhanced, and this requires an exponentially growing density of states $\hat{f}_0(E)$. More details are presented in \cite{Pappadopulo:2012jk}. Here we just emphasize that the density of states of a $p$-dimensional defect grows like the one of a $d$-dimensional CFT. This is in accordance, for instance, with the trivial defect, where the defect states coincide with the bulk ones. The comparison with the power-law behavior of eq. \eqref{intdefdens} implies that the squared OPE coefficients in eq. \eqref{weightedf} are exponentially suppressed.

Still following \cite{Pappadopulo:2012jk}, it is not hard to derive from eq. \eqref{intdefdens} a bound on the convergence of the tail of the OPE expansion, defined as
\beq
\mathcal{L}(E_*,\beta)=\int_{E_*}^\infty dE \hat{f}(E) e^{-E\beta}~.
\label{tailOPE}
\eeq
We refer the reader to \cite{Pappadopulo:2012jk} for the details of the proof, and we only report the most relevant result:
\begin{equation}
\mathcal{L}(E_*,\beta) \lesssim \frac{1}{\Gamma(2\Delta+1)} E_*^{2\D}
e^{-E_*\beta}~,\qquad E_*\gg \Delta /\beta~,\quad E_*\geq E_\textup{HL}~.\label{tailestdef}
\end{equation}
We learn that the defect OPE converges exponentially fast.
In eq. \eqref{tailestdef}, $E_\textup{HL}$ is the energy such that the Hardy-Littlewood asymptotics \eqref{intdefdens} starts being valid. $E_\textup{HL}\sim 1/\beta_0$, where in turns at $\beta_0$ the asymptotics \eqref{betaIdentity} starts being valid. The value of $\beta_0$ depends on the next operator acquiring a vev in the bulk channel OPE. 

Eq. \eqref{tailestdef} estimates the contribution of scaling operators with dimension above a certain large threshold. It is not hard to apply this estimate to the conformal block decomposition. The conformal block of a primary of dimension $\D_*$ resums the contribution of infinitely many operators with larger dimension. All these contributions are positive when $\hat{\eta}=1$. Therefore, If we define the tail of the conformal block decomposition in the notation of eq. \eqref{GtoGcal},
\beq
G_{\hat{\D}\geq\hat{\D}_*}(\hr,\heta) =\frac{1}{\hr^\D} \sum_{\hO_{\hat{\D}\geq\hat{\D}_*}} 
b_{\hO}^2\, \hg_{\hO}(\hr,\heta)~,
\eeq
we find that, for $\hat{\eta}=1$, 
\beq
G_{\hat{\D}\geq\hat{\D}_*}(\hr,\heta=1)
\lesssim \frac{1}{\Gamma(2\Delta+1)} \D_*^{2\D}
\hr^{\D_*}~,\quad \D_*\gg \Delta /\beta~.
\label{tailblockdef1}
\eeq
We see that the coupling of defect primaries decays exponentially with their scaling dimensions.

Finally, it is easy to extend the estimate \eqref{tailblockdef1} to kinematics with $\hat{\eta}\neq0$. Indeed, we can start from a representation of the blocks like eq. \eqref{GcalSumStates_D} and use Cauchy inequality, to obtain\footnote{Consider the vector
\begin{equation}
\ket{n, \hat{E}} = \mathcal{P}_{\hat{E}} \Ocal(1,n) \ket{\hat{0}}~, \notag
\end{equation}
where the projector $\mathcal{P}_{\hat{E}}$ was defined in footnote \ref{foot:projector}, and the following chain of inequalities:
\begin{equation}
\left|\braket{n, \hat{E} | n', \hat{E}}\right|\leq \sqrt{\braket{n, \hat{E} | n, \hat{E}}\braket{n', \hat{E} | n', \hat{E}}} = \braket{n, \hat{E} | n, \hat{E}}.  \notag
\end{equation}
From this eq. \eqref{cauchydefblock} follows.
}
\beq
\left| \hg_{\hO}(\hr,\heta) \right| \leq \hg_{\hO}(\hr,\heta=1)~.
\label{cauchydefblock}
\eeq
It follows that the estimate \eqref{tailblockdef1} holds for any value of the cross-ratios within the radius of convergence of the defect OPE. It is interesting to compare eq. \eqref{tailblockdef1} with the defect OPE decomposition of the two-point function of the trivial defect. The latter is just an ordinary two-point function of a scalar primary in a translational invariant CFT. In other words, the only operator appearing in the bulk block decomposition is the identity. The comparison suggests that, similarly to the estimate of \cite{Pappadopulo:2012jk}, the exponential convergence rate $\hat{r}^{\Delta_*}$ is likely to be optimal. For instance, taking $p=2,\,q=3$ and $\D=1.6$, the tail of the trivial defect is well fitted, as a function of $\D_*$, by $C(\hr) \D_*^{\gamma(\hr)} \hr^{\D_*}$, with $\gamma(\hr) \sim 1.0 \div 1.2$.

Finally, let us contrast the expansion in powers of $\hr$ and $\chi$. By inverting eq. \eqref{chitohr}, we notice two facts. The function $\hr(\chi)$ is regular when $-1<\chi<1$, and all the coefficients in the Taylor expansion around $\chi=0$ are positive. Therefore the region of convergence of the expansion in powers of $\chi$ coincides with the one in $\hr$. Furthermore, the Hardy-Littlewood theorem applies at $\heta=\cos\phi=1$, and the only difference with respect to the previous discussion is the strength of the singularity in the bulk channel. Dubbing $\beta_\chi=-\log\chi,$ we get in this case
\beq
\braket{\Ocal\Ocal}_D \overset{\beta_\chi\to 0}{\sim} (2\beta_\chi)^{-\Delta}~.
\label{betaIdentitychi}
\eeq
The tail of the OPE expansion, define as in eq. \eqref{tailOPE} with the replacement $\beta\to\beta_\chi$, is bounded by the following asymptotics
\beq
\mathcal{L}_{\chi}(E_*,\beta_\chi) \lesssim \frac{2^{-\D}}{\Gamma(\Delta+1)} E_*^{\D}
e^{-E_*\beta_\chi}~,\quad E_*\gg \Delta /\beta_{\chi}~.
\label{tailchi}
\eeq
Comparing eqs. \eqref{tailchi} and \eqref{tailestdef}, we conclude that the convergence is faster in the $\hr$ variable. Indeed, for every choice of kinematics inside the region of convergence \eqref{regDhat}, $\hr<\chi$, \emph{i.e.} $\beta_{\hr}>\beta_\chi$. For instance, the usual choice in the numerical defect bootstrap is $\xi=1$ \cite{arXiv:1210.4258,arXiv:1502.07217}, which corresponds to $\chi=1/3\simeq 0.33$ and $\hr=3-2\sqrt{2}\simeq 0.17$.

The results of this subsection do not extend trivially to the bulk channel, due to the lack of a positive configuration. The radial expansion converges absolutely, but the singularity of the sum of the absolute values in the crossed-channel is not known, and so cannot be used to give an estimate.

\subsection{The bulk channel block in the $\xi$-expansion and the Landau singularities
}
\label{subsec:landau}

It is instructive to compare the $\rho, \hrho$ coordinates with the analogous properties of other pairs of cross-ratios. As for the defect channel, we already discussed the region and rate of convergence of the expansion in $\chi$. In this subsection, we concentrate on the bulk OPE and we study the radius of convergence of the expansion of a bulk block in powers of $\xi$, defined in \eqref{crossxizeta}. We again focus on the case of two identical external scalars.

Up to an overall power $\xi^{\D/2}$, the bulk channel block has a series in $\xi$ with positive integer powers, as it can be readily established by inspection of the OPE. The radius of convergence of such a series equals the distance of the first singularity from the origin. The easiest way to discover its location is by means of the relation \eqref{xitor} to the $\rho$ coordinate. Since the $\rho$ expansion is non singular in the interior of the Euclidean region, we expect additional singularities to only come from the change of variables itself.  The inverse of eq. \eqref{xitor} is
\beq
r^2=\frac{2}{ \xi  }\left[1-\sqrt{\left(1-\zeta\xi\right)\left(1-(\zeta-1)\xi\right)}\right]  +1-2 \zeta ~.
\eeq
The change of variables has branch points at $\xi=1/\zeta$ and $\xi=1/(\zeta-1)$.
 Going around the singularities, $r^2$ is sent to $1/r^2$. 
$\xi=1/\zeta$ corresponds to $r^2=1$, while $\xi=1/(\zeta-1)$ lies outside the Euclidean region, and is mapped to $r^2=-1$. However, the second singularity limits the radius of convergence in the Euclidean region as well. More precisely, at fixed $\zeta\in [0,1]$ the $\xi$ expansion converges in the disk
\beq
\left| \xi \right| < \min \left(\frac{1}{\zeta},\frac{1}{1-\zeta}\right)~,
\label{xidiskbulk}
\eeq
which is strictly included in the region of convergence of the bulk OPE, as shown in fig. \ref{fig:xizetadomain}. 

\begin{figure}[h]
\centering
\includegraphics[scale=0.9]{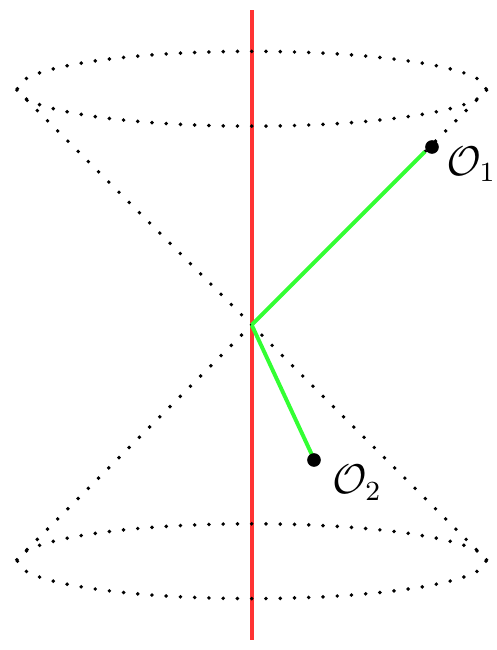}
\caption{One of the kinematics corresponding to eq. \eqref{landauconf}, in particular $\xi=1/(\zeta-1)$. The time-like defect is drawn in red. The operators $\Ocal_1$ and $\Ocal_2$ are light-like separated from the same point on the defect, as demonstrated by the light rays sketched in green. A massless particle with positive energy could be emitted by $\Ocal_2$, bounce off the defect and be adsorbed by $\Ocal_1$.}
\label{fig:landau}
\end{figure}

It is interesting to ask what is the kinematics responsible for the singularity at $\xi = 1/(\zeta-1)$. The answer lies in the study of the position space Landau diagrams \cite{Maldacena:2015iua} for correlation functions in the presence of a flat defect. The analysis of the possible singularities in perturbation theory proceeds as in \cite{Maldacena:2015iua}, except that the relevant interaction vertices are restricted to lie on the defect. This turns into the following condition: singularities can arise when a point on the defect can exchange positive energy massless particles with a subset of the external points, in a way that preserves the momentum parallel to the defect. In the case of a two-point function, this requires the external insertions to be light-like separated from the same point on the defect. Momentum conservation then requires the interaction vertex to be aligned with the projection of the external points on the defect.
 An example of this kinematics is sketched in fig. \ref{fig:landau}. 
If these requirements are met, the cross ratios obey
\beq
\xi= \frac{1}{2}(\epsilon- \cos\phi)~,
\qquad \epsilon=\pm 1~,
\label{landauconf}
\eeq
where $\cos \phi$ is the cross ratio defined in eq. \eqref{crosschiphi} and  $\epsilon=\pm 1$
if the defect is space-like or time-like respectively (we use the mostly plus signature, and we keep the notation $\cos\phi$ even if the cross-ratio may not be bounded by one). For a space-like defect, eq. \eqref{landauconf} has the two solutions $\zeta=1$ and $\xi=0$, whose kinematics are easy to visualize but are not important here.
 Instead, $\xi=1/(\zeta-1)$ is the general solution when the defect is time-like, and the kinematics is the one depicted in fig. \ref{fig:landau}. It should be noted that, in this case, $-1\leq\xi\leq 0$ and $\zeta\leq 0$. 
  Let us now consider the path to reach this kinematics from a Euclidean configuration. When the two external operators both lie at $t=0$, $t$ being the time-like coordinate parallel to the defect, the cross-ratios are in the Euclidean region.
   If we start separating them along $t$ keeping the angle $\phi$ fixed, they become light-like separated before reaching the configuration \eqref{landauconf}. This amounts in crossing the point $(\xi=0,\zeta=\infty)$, which is only a branch point for the prefactor $\xi^{\Delta/2}$ of the block, and lies within the radius of convergence of the radial expansion. This justifies the appearance of this singularity on the first Riemann sheet in the complex $\xi$ plane, and the fact that it limits the radius of convergence of the $\xi$ expansion.

\subsubsection{The self-dual point and the bootstrap}
\label{subsec:selfdual}

We would like to end this section with some comments related to the bootstrap. In the case of a boundary CFT, the point that has been used so far for the numerical exploration is $\xi=1$ \cite{arXiv:1210.4258,arXiv:1502.07217}. The natural generalization for it is $(\xi,\zeta)=(1,0)$, or in terms of the $\r$ coordinates,
\beq
r^2=\hr=3-2\sqrt{2}\simeq 0.17~,   \qquad \eta=\heta=1.
\label{selfdual}
\eeq
One appealing feature of this value follows from eq. \eqref{rho(hrho)}. This equation is an involution for the pair of coordinates $\rho\to\sqrt{\hrho}$, meaning that it is its own inverse. The point $\hrho=\rho^2=3-2\sqrt{2}$ is the fixed point of the involution, and corresponds to eq. \eqref{selfdual}. 
One would like to motivate the choice \eqref{selfdual} quantitatively, namely as the point in which the bulk and the defect OPEs converge at the same rate, but this appears harder as a consequence of the lack of control over the convergence of the bulk channel OPE.

\begin{figure}[t]
\centering
\includegraphics[scale=0.6]{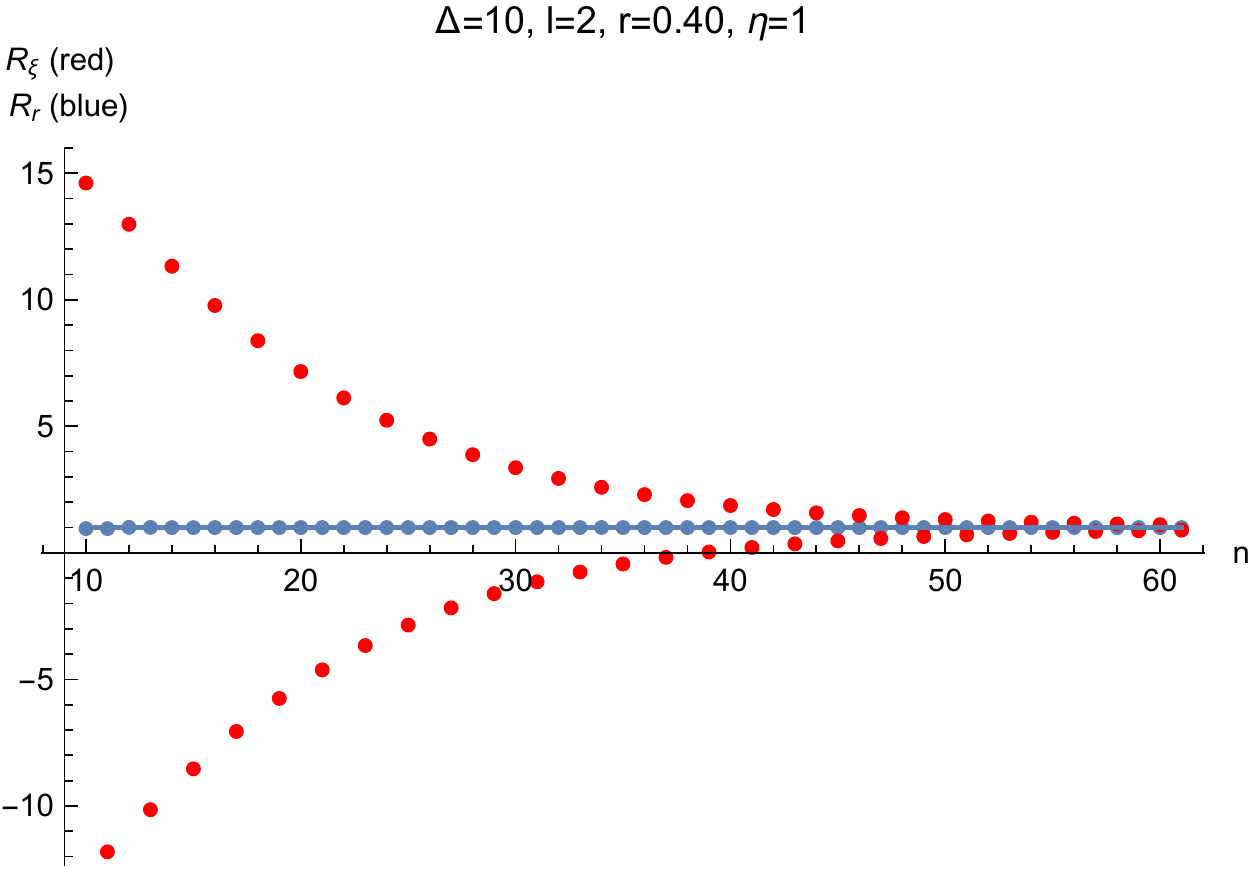}\quad
\includegraphics[scale=0.6]{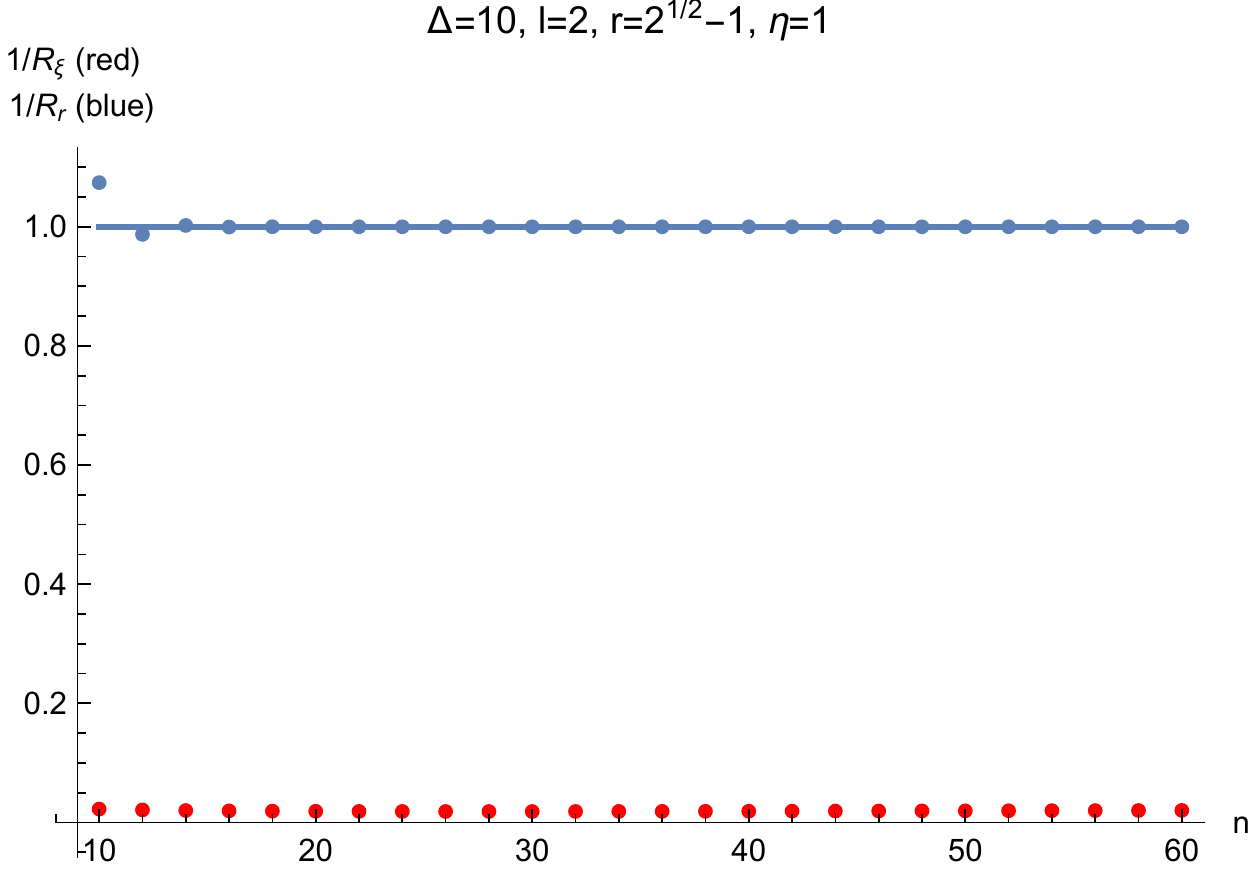}
\caption{Comparison of the expansion of the block $g_{\Delta l}$ of dimension $\Delta=10$ and spin $l=2$ in powers of $\xi$ (in red) and $r$ (in blue). The meaning of the labels is as in eq. \eqref{RrRxi}. The blue solid line is constant and equal to 1. The oscillation of $R_r$ and $R_\xi$, especially visible in the latter, follows from the fact that the two series expansions of the block are alternating. On the right, $r$ is at the self-dual point, eq. \eqref{selfdual}. We plot $R_{r,\xi}^{-1}$ because $|R_{\xi}|$ diverges at large $n$ for $\xi\geq 1$.}
\label{fig:d4q2convergence}
\end{figure}

We would like to stress that the self-dual point \eqref{selfdual}, \emph{i.e.} $(\xi,\zeta)=(1,0)$, lies at the boundary of the region of convergence of the $\xi$-expansion - see eq. \eqref{xidiskbulk}. This makes the use of the $\rho$ coordinate not only convenient but strictly necessary in the bulk channel, in the cases where the bulk blocks are not known in closed form. We exemplify this fact by comparing the convergence of the $\xi$ and $r$ expansions in the case of a block known in closed form. As we discuss in appendix \ref{app:d4q2}, the blocks for two identical scalar primaries with a codimension 2 defect in 4 dimensions belong to this category.
We simply expand the block in powers of $\xi$ and $r$ at fixed $\eta$, and 
we define the two ratios
\begin{equation}
\begin{split}
R_r(n) &= \frac{1}{g_{\D l}}\left(\textup{Series expansion of $g_{\D l}$ up to order $r^{\D+n}$}\right)~, \\
R_\xi(n) &= \frac{1}{g_{\D l}}\left(\textup{Series expansion of $g_{\D l}$ up to order $\xi^{\D/2+n}$}\right)~.
\end{split}
\label{RrRxi}
\end{equation}
We suppressed the dependence on $\eta$. We show an example in fig. \ref{fig:d4q2convergence}, for $\eta=1\, (\zeta=0)$ and two values of $r$: $r=0.4$, which corresponds to $\xi=0.91$, and $r=\sqrt{2}-1$ ($\xi=1$), \emph{i.e.} the self-dual point \eqref{selfdual}. 
It is evident in fig. \ref{fig:d4q2convergence} that the $\xi$-expansion does not converge to the value of the block at the self-dual point. Furthermore, when $r=0.4$ and both series converge, the expansion in $r$ clearly shows a faster rate of convergence: a good approximation of the block is already obtained with the inclusion of the descendants up to level $\sim 10$  when $\Delta=10$. For instance, the truncation to order $r^{\Delta+12}$ differs from the exact block by less than $0.7\%$.

\section{Conclusions}

In this work, we introduced the radial coordinates $(r,\eta)$ and $(\hr,\heta)$ for the study of a two-point function of local operators in the presence of a conformal defect. The new coordinates have numerous advantages which, much as in the case of the radial coordinates for the four-point function \cite{Hogervorst:2013sma}, largely follow from their clean geometric interpretation. In particular, the expansion of the bulk (defect) conformal blocks in powers of $r$ ($\hr$) is closely related to the Euclidean OPE, and is therefore easily organized. This is of particular importance for the computation of conformal blocks exchanged by operators with spin \cite{Lauria:2018klo}. Furthermore, the region of convergence of the bulk (defect) OPEs coincide with the region of convergence of the expansion of the two-point function in powers of $r$ ($\hr$). This signals that the latter expansions are especially well behaved, a fact that we have quantified in the defect channel by showing that the rate of convergence is exponential. While we did not estimate the rate of convergence of the bulk channel ($r$) expansion, we compared it with the expansion in the cross ratio $\xi$ -- see eq. \eqref{crossxizeta} -- which is customarily used in the defect CFT literature. We pointed out that the latter has a smaller radius of convergence, which in particular makes it unsuitable for the numerical bootstrap applications. Finally, since the radial coordinates give a strong geometric intuition on the position of the singularities of the two-point function both in Euclidean and Lorentzian signatures, we expect that they may play a role in the development of analytic approaches to the defect conformal bootstrap as well.

\section*{Acknowledgments}

We would like to thank J. Penedones for numerous useful suggestions, and in particular V. Gonçalves for collaboration in the early stages of the project and for many useful discussions. We also thank ICTP-SAIFR in S\~{a}o  Paulo for hospitality during the Bootstrap 2017 workshop, and the Simons Collaboration on the Non-perturbative Bootstrap for providing many stimulating workshops and conferences during which this work carried out. In addition, we thank the Perimeter Institute for Theoretical Physics, where part of the project was carried out.
Research at Perimeter Institute is supported by the Government of Canada through Industry Canada and by the Province of Ontario through the Ministry of Research \& Innovation.

EL would like to thank A. Bernamonti, M. Bill\`o, N. Bobev, L. Di Pietro, F. Galli and M. Hogervorst for many interesting conversations and suggestions. EL is supported by the Belgian Federal Science Policy Office through the Inter-University Attraction Pole P7/37, by the COST Action MP1210,  European Research Council grant no. ERC-2013-CoG 616732 HoloQosmos, as well as the FWO  Odysseus  grants G.001.12 and G.0.E52.14N\@.

MM  is supported by the Simons
Foundation grant 488649 (Simons collaboration on the non-perturbative bootstrap) and by
the National Centre of Competence in Research SwissMAP funded by the Swiss National Science
Foundation.

The work of ET is supported by the Simons Foundation grant 488655 (Simons Collaboration on the Nonperturbative Bootstrap), by the Portuguese Funda\c{c}\~ao para a Ci\^encia e a Tecnologia (FCT) through the fellowship SFRH/BD/51984/2012 and by Perimeter Institute for Theoretical Physics.

\appendix


\section{Normalizations}\label{app:conventions}

Here we report our choice of normalizations of the CFT data, through their appearance in the correlation functions. All the operators are canonically normalized, except the defect primaries with non vanishing transverse spin $s$, whose normalization is in eq. \eqref{twopoint_D}.

It is now standard to use the technology of \cite{SpinningCC} to write correlation functions of traceless symmetric operators.
To each operator $\Ocal_i(P_i,Z_i)$ is associated a polarization vector $Z_i$ which contracts its indices. $Z_i$ is orthogonal to $P_i$, \emph{i.e.} $Z_i\cdot P_i=0$, and null, \emph{i.e.} $Z_i\cdot Z_i=0.$ The projection of $Z_i$ on the Poincaré section \eqref{PoincareSec} defines polarization vectors $z_i$ in physical space according to:
\be
Z=(z\cdot x, z^\m, -z\cdot x).
\label{Zpoin}
\ee

As for the correlation functions of local operators without defects, for the purposes of this paper we need the spin 0 - spin 0 - spin $l$ three-point function, which are fixed by conformal invariance up to a coefficient:
 \be
 \langle 
\Ocal_1(P_1)\Ocal_2(P_2) \Ocal_3(P_3,Z_3)
\rangle
= c_{123} 
\frac{(V_{3,12})^l}{(P_{12})^{\alpha_{123}}(P_{13})^{\alpha_{132}}(P_{23})^{\alpha_{231}}},
\label{three_point_BBB}
\ee
where we defined $P_{ij}\equiv (-2 P_i \cdot P_j)$ and $\alpha_{ijk}\equiv (\D_i+\D_j-\D_k)/{2}$ and
\be\label{three_point_BBB0}
V_{i,jk} 
= \dfrac{(P_i\cdot P_j) (Z_i\cdot P_{k}) - (P_i\cdot P_{k}) (Z_i\cdot P_j)}{\sqrt{- 2 (P_i\cdot P_j)( P_j\cdot P_k)( P_k\cdot P_i)}} \ .
\ee

Turning to the correlation functions in the presence of a defect, the
one-point functions of symmetric and traceless bulk primaries with even spin $l$ are \cite{Billo:2016cpy}
\begin{align}
 \langle 
\Ocal(P_1,Z_1)
\rangle_D
= a_{\Ocal}\,
\frac{Q_l}{(P_1 \tdot P_1)^{\frac{\D}{2}}} \ , \quad Q_l= \left( \frac{(P\tdot Z)^2}{(P \tdot P)}-Z \tdot Z\right)^{l/2}.
\label{onepoint}
\end{align}
The two point functions of defect operators of transverse spin $s$ and parallel spin $\hl=0$ are normalized as follows:
 \be
 \label{twopoint_D}
 \langle 
\hat \Ocal(P_1,W_1)
\hat \Ocal(P_2,W_2)
\rangle_D
= \frac{(W_1\wbullet W_2)^s }{(-2 P_1 \pdot P_2)^{\frac{\hD}{2}}} \ .
\ee
Here $W_i$ is a polarization vector that contracts the transverse spin indices of $\hO_i$.
Finally, the two point function of a scalar bulk primary and  a defect primary with transverse spin $s$ and vanishing parallel spin is given by
\begin{align}\label{btodefect2pt}
\langle \Ocal(P_1)  \hO(P_2,W_2) \rangle_D = b_{\Ocal\hO}\,\frac{(Q_{BD}^1)^s}{(-2 P_1 \pdot P_2)^{\hD} (P_1 \tdot P_1)^{\frac{\D-\hD}{2}}},
 \quad Q_{BD}^1 \equiv \frac{P_1 \wbullet W_2}{(P_1 \wbullet P_1)^{1/2}}.
\end{align}


\section{Details on the bulk channel scalar conformal block}
\label{app:scalarBulk}
\subsection{Casimir equation}
In section \ref{sec:scalar} we explained that the conformal partial waves are eigenfunctions of the conformal Casimir.
Using \eqref{GtoGcal}, it is not hard to rewrite the Casimir equation \eqref{Casimir_Scalar_Bulk} as a differential equation for the conformal block $g_{\D,l}$,
\ba
\label{CasimirScalarBulk}
\!\!\!\!\!\!\!\!\!
\begin{split}
&\frac{1}{2} \left(r^4-1\right)  \left(f^2 c_{\Delta ,l}+4 \Delta _{12}^2 r^2 \left(2 \eta ^2 \left(r^4+1\right)-\left(r^2+1\right)^2\right)\right)g_{\D,l}\\
&-f \left(r^4-1\right) \left(f d \left(\eta ^2-1\right)-4 \eta ^2 r^2 \left(2 \eta ^2+q-3\right)+(q-1) \left(r^2+1\right)^2\right)  \partial_{\eta^2}  g_{\D,l} \\
&-f  \left(f d \left(r^2-1\right)^2-16 \eta ^2 r^2 \left((q-1) r^2-1\right)+2 \left(r^2+1\right)^2 \left(2 (q-2) r^2+r^4-1\right)\right)r^2 \partial_{r^2} g_{\D,l}\\
&-2 f^2 \eta ^2 \left(\eta ^2-1\right) \left(r^4-1\right) \partial_{\eta^2}^2 g_{\D,l}
-2 f^2 \left(r^4-1\right) r^4  \partial_{r^2}^2g_{\D,l}=0 \ , \\
\end{split}
\ea
where we defined the auxiliary function $f=(r^2+1)^2-4 \eta^2r^2$.

In the following solve eq. \eqref{CasimirScalarBulk} at leading order when $r$ is small and at leading order when $\D$ is large.
In both cases it is useful to define a new function $h_{\D,l}$, obtained by stripping out a factor $(4r)^{\D} $ in front of the conformal blocks -- see also eq. \eqref{LeadingOPEBulkScalarCB}:
\be
h_{\D,l}(r,\eta) \equiv  (4r)^{-\D} g_{\D,l}(r,\eta ) \ .
\ee
At leading order in $r$, \eqref{CasimirScalarBulk} becomes
\be
\label{CasimirBulkLeadingrScalar}
 l (d+l-2) h_{\D,l}(0,\eta) -2   \left(d \, \eta ^2-p-1\right) \partial_{\eta^2} h_{\D,l}(0,\eta) -4 \left(\eta ^2-1\right) \eta ^2 \partial_{\eta^2}^2 h_{\D,l}(0,\eta)=0 \,.
\ee
The functions \eqref{DEF:Cj}, with $j=l$, provide a solution to this equation compatible with the OPE limit, as explained in subsection \ref{BulkNaturalExpansionScalar}.

Similarly, at leading order at large $\D$ the Casimir equation reduces to
\be
 \left(d \left(f-f r^2\right)-2 f q-4 \left(\eta ^2-1\right) \left(r^2+1\right)^2\right)h_{\infty l} -2 f \left(r^4-1\right) \partial_{r^2}h_{\infty l}=0 \ .
\ee
This first order differential equation can be solved explicitly and the boundary condition is again provided by the leading OPE limit \eqref{LeadingOPEBulkScalarCB}.
The full result is
\be
\label{BulkLargeDeltaScalar}
h_{\infty l}(r,\eta)= \frac{ \left(1-r^2\right)^{1-\frac{q}{2}} \left(r^2+1\right)^{\frac{q-d}{2}}}{\sqrt{\left(r^2+1\right)^2-4 \eta ^2 r^2}} \mathfrak{C}_l(\eta) \ .
\ee

\subsection{The Zamolodchikov Recurrence Relation }
\label{App:ZamolodchikovBulkScalar}
In this section we explain how to compute all the ingredients for the Zamolodchikov recurrence relations presented in section \ref{BulkZamolodchikovScalar}.
Most of the ingredients were already computed in \cite{recrel}, so we will briefly review how the computation was done. 
Finally we explain how to compute the new ingredient $M_A^{(L)}$ defined in \eqref{MbulkSca}.

For generic values of $p$ and $q$, only bulk operators transforming in symmetric and traceless representations of even spin $l$ are allowed to appear in the OPE of two bulk scalar operators.
For symmetric and traceless representations, all the null states were found explicitly in \cite{recrel}. In the following we review how to generate them.
Consider a traceless and symmetric primary state with spin $l$:
\be
|\D,l \,;z \rangle \equiv z_{\m_1} \dots z_{\m_l} \Ocal^{\mu_1 \dots \mu_l}(0)|0\rangle\ \equiv \Ocal(z,0)|0\rangle\ .
\ee
Following the notation of \cite{recrel}, all the primary descendant states can be written as a differential operator $\mathcal{D}_A$ acting on the primary state:
\be
| \D_A,l_A\,; z \rangle = \Dcal_A  |\D,l \,;z \rangle 
\label{DA}
\ee
where $A=T,n$ with $T=\I, \II, \III$ and $n=1,2 \dots$. 

\be
\label{OA_def}
\begin{array}{ll}
\Dcal_{\I,n}|\D,l\, ;z \rangle
&\equiv (z\cdot P)^n |\D,l\, ;z \rangle \ ,
\\
\Dcal_{\II,n}|\D,l\, ;z \rangle 
&\equiv
\frac{(D_z \cdot P)^n}{(2-d/2-l)_n(-l)_n} |\D,l\, ;z \rangle \ ,
\\
\Dcal_{\III,n}|\D,l\, ;z \rangle 
&
\equiv  \Vcal_0 \cdot \Vcal_1 \cdots \Vcal_{n-1}
|\D,l\, ;z \rangle \ , 
\end{array}
\ee
where 
\be  \label{Vcalj}
\Vcal_j\equiv P^2-2 \frac{ (P\cdot z) (P\cdot D_z)}{ (d/2+l+j-1) (d/2+l-j-2)}  \ .
\ee
The state $| \D_A,l_A\,; z \rangle$ becomes primary when $\D=\D^\star_A$ defined by
\be
\label{Dstar}
\begin{array}{l c l l}
\D^\star_{\I, n} &\equiv& 1-l-n   & \qquad\qquad n=1,2,\dots \ , \\
\D^\star_{\II, n}&\equiv& l+d-1-n  & \qquad\qquad n=1,2,\dots, l \ , \\
\D^\star_{\III, n}&\equiv& \frac{d}{2}-n & \qquad\qquad n=1,2, \dots \ .
\end{array}
\ee

The quantity $Q_A$ is defined as the inverse of the norm of the states \eqref{OA_def} at the pole $\D=\D^\star_A$
\be
\langle \Ocal_A\Ocal_A\rangle ^{-1} = 
\frac{Q_A}{\D- \D^\star_A} +O((\D- \D^\star_A)^0)
\ .
\ee
It was computed in \cite{recrel}:
\begin{align}
\label{QA}
\begin{split}
&Q_{\I,n}= -\frac{n }{2^n (n!)^2} \ , \\
&Q_{\II,n} =  -\frac{n (-l)_n}{(-2)^n (n!)^2 (d+l-n-2)_n} \frac{(d/2+ l- n-1)}{ (d/2+l-1)} \ , \\
&Q_{\III,n} =-\frac{n }{(-16)^n (n!)^2 (d/2-n-1)_{2 n}} \frac{(d/2+l-n-1)}{ (d/2+l+n-1)} \ .
\end{split}
\end{align}
In \cite{recrel} also $M_A^{(R)}$ was obtained, which appears as a normalization coefficient of the three point function with a primary descendant operator
$\langle \Ocal_1\Ocal_2\Ocal_A\rangle=M_A^{(R)} \langle \Ocal_1\Ocal_2\Ocal\rangle$. 
This quantity can be obtained by performing the following computation
\be
\Dcal_A \frac{(-x\cdot z)^l}
{(x^2)^\frac{\D+\D_{12}+l}{2}} = M_A^{(R)}   \frac{(-x\cdot z)^{ l_A}}
{(x^2)^\frac{\D_A+\D_{12}+ l_A}{2}},
\label{DA-MA}
\ee
to find
\begin{align}
\begin{split}
\label{Mscalar}
M^{(R)}_{\I, n}&= 
(2i)^n \left(\frac{\D+\D_{12}+l}{2}\right)_n \ ,\\
M^{(R)}_{\II, n}&= i^n \frac{(d+l-n-2)_n }{\left(d/2+l-n-1\right)_n} \left(\frac{\Delta+ \Delta_{12}+2-d-l}{2}  \right)_n \ ,\\
M^{(R)}_{\III, n}&=
(-4)^n \frac{(d/2-n-1)_l}{(d/2+n-1)_l}  \left( \frac{\Delta +\Delta_{12}+2-d-l}{2}\right) _n \left( \frac{l+\Delta +\Delta_{12}}{2}\right) _n \ .
\end{split}
\end{align}
In the computation of the residue $R_A$ in eq. \eqref{QMM} $M^{(R)}_{A}$ should be evaluated at $\D=\D^\star_A$.

Similar manipulations lead to the results for $M^{(L)}_A$. We consider the one point function \eqref{onepoint} projected onto the Poincar\'e section:
\be
\langle
\Ocal_{\D \, l}(x,z)
\rangle_D
\equiv 
a_\Ocal \;
\frac{\left(\left(x\tdot z\right)^2-(z\tdot z) (x\tdot x)\right){}^{\frac{l}{2}}}{\left(x \tdot x\right){}^{\frac{\Delta+l}{2}}}  \ ,
\ee
where $\tdot$ is the scalar product for a flat defect defined in eq. \eqref{tdot}.
$M_A^{(L)}$ is defined by the equation
\be
\Dcal_A 
 \langle \Ocal_{\D^\star_A \, l}(x,z)\rangle_D =M_A^{(L)} \langle \Ocal_{\D_A\, l_A}(x,z)\rangle_D   \ .
\ee
The result is
\begin{align}
\label{MRBulk}
\begin{split}
M_{\I,n}^{(L)}&=((n-1)\text{!!})^2 \ , \\
M_{\II,n}^{(L)}&=((n-1)\text{!!})^2 \ \frac{ \left(\frac{-d-l+4}{2}\right)_{\frac{n}{2}} \left(\frac{-l-q+3}{2}\right)_{\frac{n}{2}} \left(\frac{-d-l+q+1}{2} \right)_{\frac{n}{2}}}{\left(\frac{1-l}{2}\right)_{\frac{n}{2}} \left(\frac{-d-2 l+4}{2}\right)_n} \ , \\
M_{\III,n}^{(L)}&=(-4)^n \frac{\left(\frac{d/2-q-n+2}{2}\right)_{n}\left(\frac{d/2+l-n}{2}\right)_{n}\left(\frac{d/2-n-1}{2}\right)_{n}}{\left(\frac{d/2-n+l-1}{2}\right)_{n}} \ .
\end{split}
\end{align}
\\

\section{The scalar bulk blocks in $d=4$, $q=2$}
\label{app:d4q2}

\begin{figure}[ht]
\centering
\includegraphics[scale=0.8]{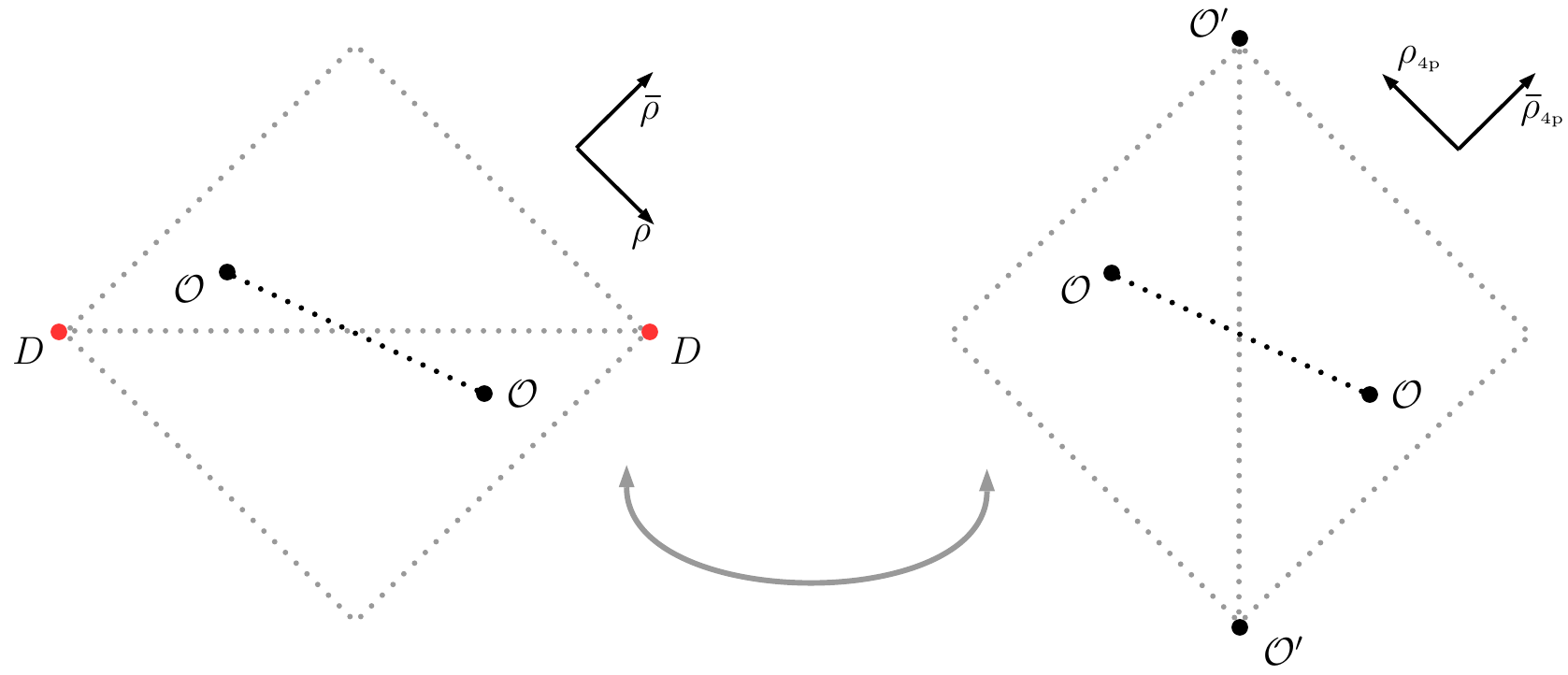}
\caption{The geometric equivalence of a codimension two defect to a pair of local operators. In Minkowsky space, $(\rho,\bar{\rho})$ are independent real coordinates. The locus of the defect on the left -- in red -- is the intersection of the light-cones of the two time-like separated points at the top and bottom of the diamond on the right.}
\label{fig:rho4rho}
\end{figure}

The conformal blocks for two identical external primaries with a codimension 2 defect are the same as the blocks of an ordinary four-point function of pairwise identical operators \cite{Billo:2016cpy}. In particular, the bulk channel blocks for external scalars are known exaclty for a two-dimensional surface defect in four dimensions. They are most conveniently written in terms of auxiliary variables $z,\,\bar{z}$:
\begin{multline}
g_{\D l}(r,\eta) \propto \frac{z \bar{z}}{z-\bar{z}} \left(k_{\D+l}(z)k_{\D-l-2}(\bar{z})-z\leftrightarrow \bar{z}\right), \qquad
k_\beta(z)=z^{\beta/2} {}_2F_1(\beta/2,\beta/2,\beta,z).
\label{blockd4q2}
\end{multline}
We do not pay attention to the normalization in this appendix.
The $(z,\bar{z})$ variables obey
\beq
z\bar{z} =- \xi e^{-\ii \phi}, \quad (1-z)(1-\bar{z}) = e^{-2\ii \phi}.
\label{zxi}
\eeq
This translates to
\beq
z=\frac{\hrho-1}{\hrho}~,\quad \bar{z}=1-\bar{\hrho}~,
\eeq
or in terms of the bulk-channel coordinates,
\beq
z=-\frac{4\r}{(1-\r)^2}~,\quad \bar{z}=\frac{4\bar{\r}}{(1+\bar{\r})^2}~.
\eeq
The last relation is especially simple when re-expressed in terms of the $\rho$-coordinate for the four-point function, which we rename $\rho_\textup{4p}$:
\beq
\rho_\textup{4p}=-\rho~,\qquad \bar{\rho}_\textup{4p}=\bar{\r}~.
\label{rho4rho}
\eeq
The geometric interpretation of equation \eqref{rho4rho} is simple, and depicted in fig. \ref{fig:rho4rho}. 
In figure on the right we show the four-point function $\langle \Ocal(x_1)\Ocal(x_2) \Ocal'(x_3) \Ocal'(x_4)\rangle$ in the radial frame in Lorentzian signature. The conformal transformations that leave the positions of $\Ocal'(x_3)$ and $\Ocal'(x_4)$ invariant also leave invariant their future and past lightcones. The intersection of the lightcones is a codimension 2 sphere, so we conclude that a pair of points and a sphere have the same stabilizer. 

 The change of variables \eqref{zxi} was first found in  \cite{Billo:2016cpy}, and then given a geometric interpretation in \cite{Gadde:2016fbj}.  Later, the correspondence between pairs of local operators and a class of codimension two surfaces re-emerged in \cite{Czech:2016xec,deBoer:2016pqk}. 
We stress that this is a kinematic correspondence rather than a duality. In fact the reasoning offers no evidence that correlation functions should agree. 

  That said, the availability of exactly known conformal blocks is welcome.
For instance, from the closed expression \eqref{blockd4q2} we can check that the blocks are regular in the $\mathcal{D}$ or $\hDcal$ domains in eqs. \eqref{regD} and \eqref{regDhat}. In particular, the cuts of the hypergeometric functions in eq. \eqref{blockd4q2} are mapped to the circles $|\rho|=1$ and $|\bar{\rho}|=1$, or equivalently to the negative real axes of $\hrho$ and $\bar{\hrho}$. More precisely, the line $\hrho=\bar{\hrho}\in [-1,0]$ is mapped to $\bar{z}=z/(z-1)$ with $z\in [2,\infty)$. Hence, the bulk channel conformal block is discontinuous when crossing the negative real axis in the $\hrho$ plane.


\vspace{30pt}


%

\end{document}

%% file: Embedding1.pdf_tex
\begingroup%
  \makeatletter%
  \providecommand\color[2][]{%
    \errmessage{(Inkscape) Color is used for the text in Inkscape, but the package 'color.sty' is not loaded}%
    \renewcommand\color[2][]{}%
  }%
  \providecommand\transparent[1]{%
    \errmessage{(Inkscape) Transparency is used (non-zero) for the text in Inkscape, but the package 'transparent.sty' is not loaded}%
    \renewcommand\transparent[1]{}%
  }%
  \providecommand\rotatebox[2]{#2}%
  \ifx\svgwidth\undefined%
    \setlength{\unitlength}{1008.71386719bp}%
    \ifx\svgscale\undefined%
      \relax%
    \else%
      \setlength{\unitlength}{\unitlength * \real{\svgscale}}%
    \fi%
  \else%
    \setlength{\unitlength}{\svgwidth}%
  \fi%
  \global\let\svgwidth\undefined%
  \global\let\svgscale\undefined%
  \makeatother%
\begin{picture}(1,0.4)%
 \put(0,0){\includegraphics[scale=0.5]{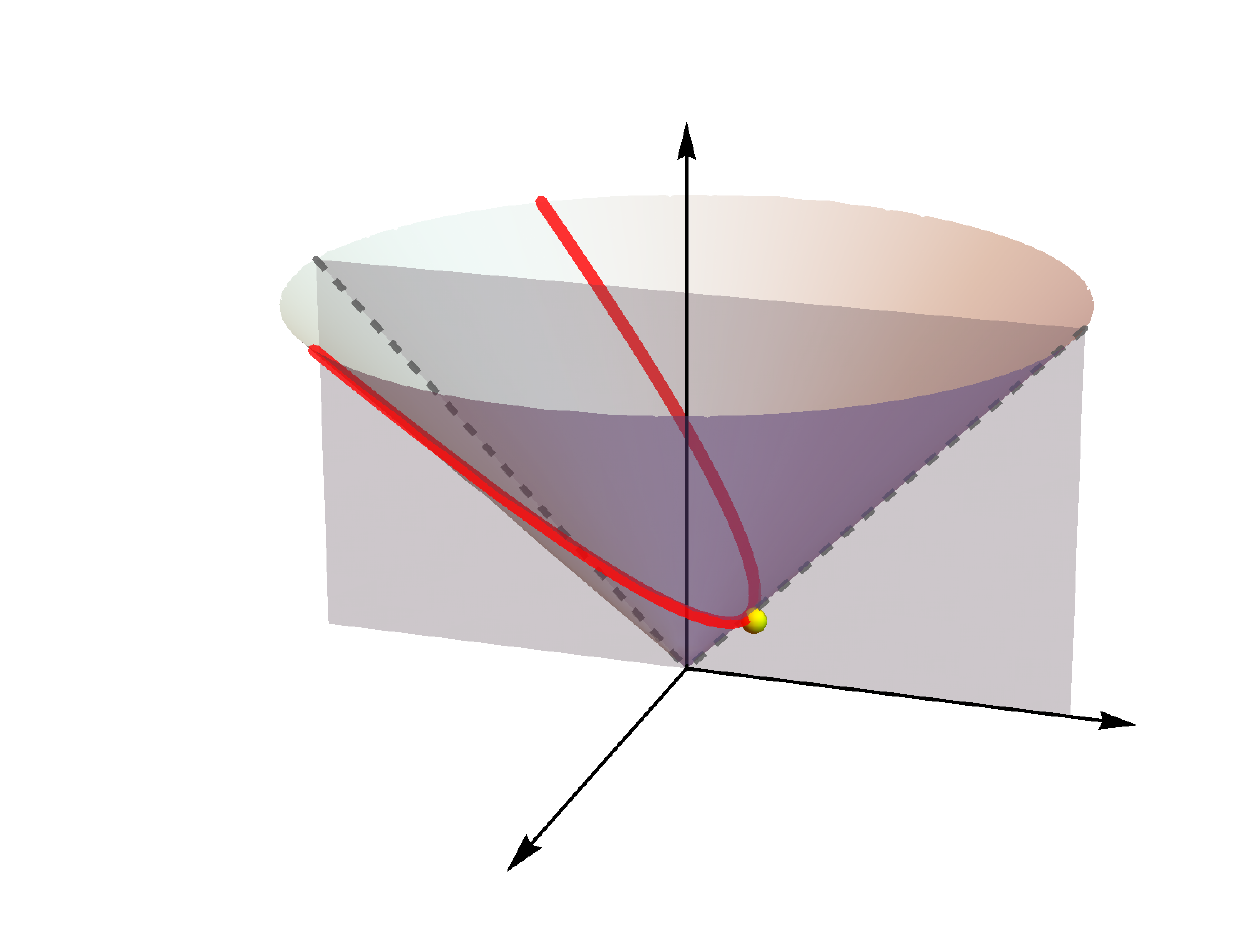}}
\put(0.81,0.12){\color[rgb]{0,0,0}{\makebox(0,0)[lb]{\scriptsize $P^2$}}}
\put(0.31 ,0.07){\color[rgb]{0,0,0}{\makebox(0,0)[lb]{\scriptsize $P^1$}}}
\put(0.51,0.6){\color[rgb]{0,0,0}{\makebox(0,0)[lb]{\scriptsize $P^0$}}}
  \end{picture}%
\endgroup%

%% file: Embedding2.pdf_tex
\begingroup%
  \makeatletter%
  \providecommand\color[2][]{%
    \errmessage{(Inkscape) Color is used for the text in Inkscape, but the package 'color.sty' is not loaded}%
    \renewcommand\color[2][]{}%
  }%
  \providecommand\transparent[1]{%
    \errmessage{(Inkscape) Transparency is used (non-zero) for the text in Inkscape, but the package 'transparent.sty' is not loaded}%
    \renewcommand\transparent[1]{}%
  }%
  \providecommand\rotatebox[2]{#2}%
  \ifx\svgwidth\undefined%
    \setlength{\unitlength}{1008.71386719bp}%
    \ifx\svgscale\undefined%
      \relax%
    \else%
      \setlength{\unitlength}{\unitlength * \real{\svgscale}}%
    \fi%
  \else%
    \setlength{\unitlength}{\svgwidth}%
  \fi%
  \global\let\svgwidth\undefined%
  \global\let\svgscale\undefined%
  \makeatother%
  \begin{picture}(1,0.1)%
 \put(0,0){\includegraphics[scale=0.5]{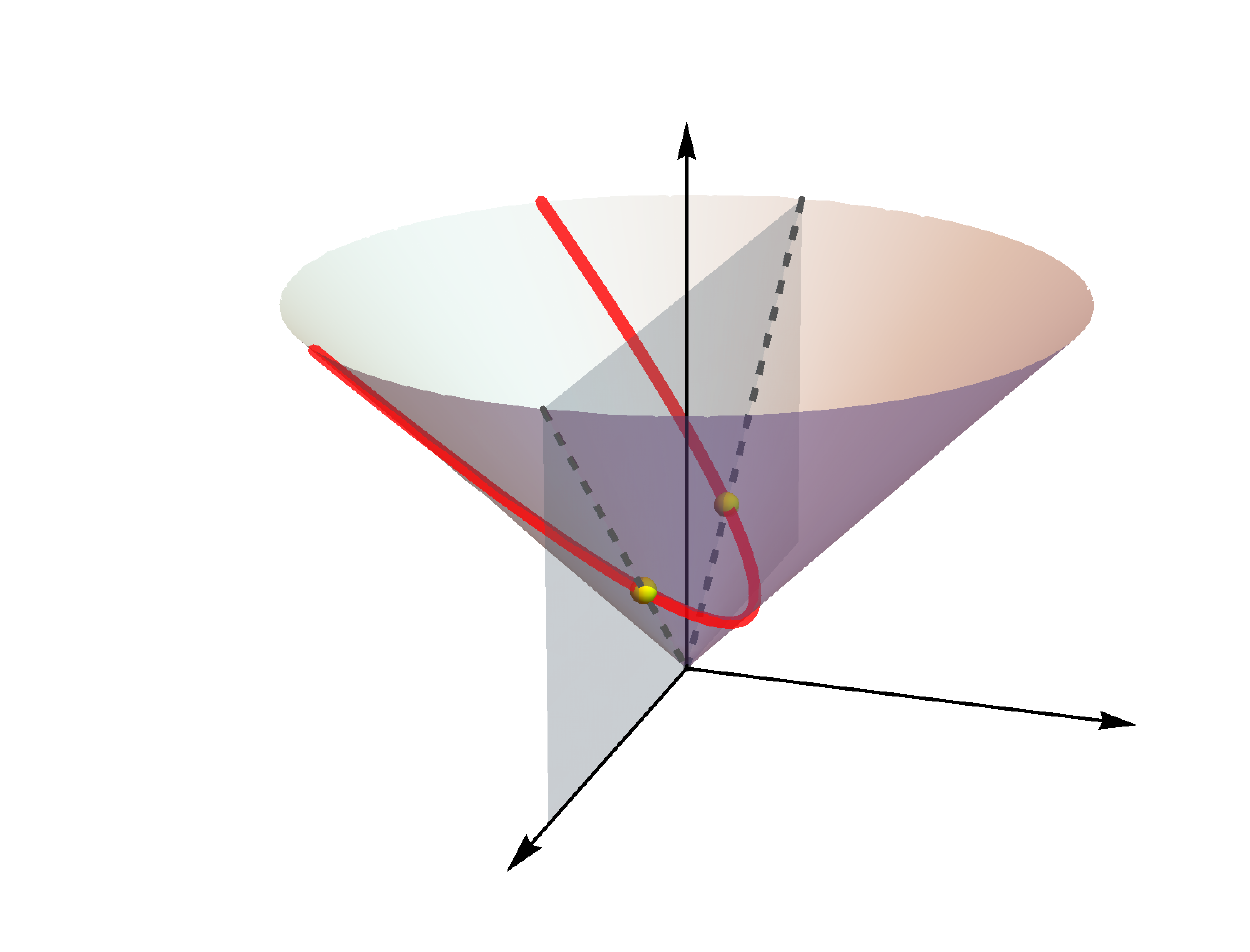}}
\put(0.062,0.014){\color[rgb]{0,0,0}{\makebox(0,0)[lb]{\scriptsize $P^1$}}}
\put(0.16 ,0.023){\color[rgb]{0,0,0}{\makebox(0,0)[lb]{\scriptsize $P^2$}}}
\put(0.1,0.117){\color[rgb]{0,0,0}{\makebox(0,0)[lb]{\scriptsize $P^0$}}}
  \end{picture}%
\endgroup%

%% file: scalarrec.pdf_tex
\begingroup%
  \makeatletter%
  \providecommand\color[2][]{%
    \errmessage{(Inkscape) Color is used for the text in Inkscape, but the package 'color.sty' is not loaded}%
    \renewcommand\color[2][]{}%
  }%
  \providecommand\transparent[1]{%
    \errmessage{(Inkscape) Transparency is used (non-zero) for the text in Inkscape, but the package 'transparent.sty' is not loaded}%
    \renewcommand\transparent[1]{}%
  }%
  \providecommand\rotatebox[2]{#2}%
  \ifx\svgwidth\undefined%
    \setlength{\unitlength}{1008.71386719bp}%
    \ifx\svgscale\undefined%
      \relax%
    \else%
      \setlength{\unitlength}{\unitlength * \real{\svgscale}}%
    \fi%
  \else%
    \setlength{\unitlength}{\svgwidth}%
  \fi%
  \global\let\svgwidth\undefined%
  \global\let\svgscale\undefined%
  \makeatother%
  \begin{picture}(1,1)%
    \put(0,0){\includegraphics[width= \unitlength]{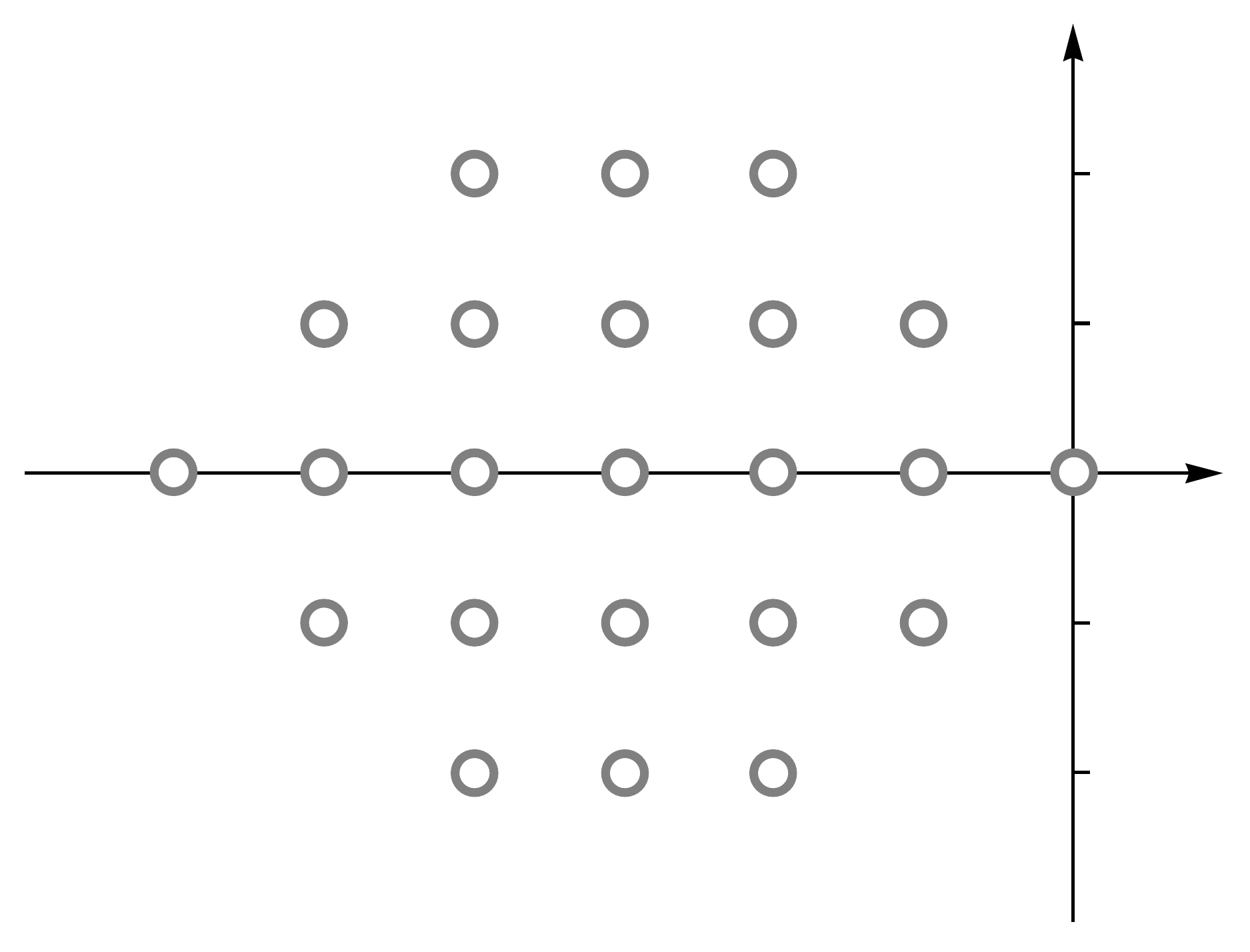}}%

\put(0.93 ,0.32){\color[rgb]{0,0,0}{\makebox(0,0)[lb]{$n$}}}
\put(0.9 ,0.7){\color[rgb]{0,0,0}{\makebox(0,0)[lb]{$k$}}}
    
  \end{picture}%
\endgroup%